\documentclass[12pt]{article}

\usepackage{cite}
\usepackage{subfigure}
\usepackage{multirow}
\usepackage{helvet}
\usepackage{amsmath}
\usepackage{amssymb}
\usepackage{setspace}
\usepackage{setspace}
\usepackage{graphicx}
\usepackage{empheq}
\usepackage{subfigure}

\def\be{\begin{equation}}
\def\ee{\end{equation}}

\begin{document}

\titlepage                                                    

\begin{flushright}
IPPP/17/85 \\
\today \\
\end{flushright}

 \vspace*{0.5cm}
\begin{center}                                                    
{\Large \bf The Impact of LHC Jet Data on the MMHT\\ \vspace{0.4cm}PDF Fit at NNLO}\\

\vspace*{1cm}
                                                   
L. A. Harland--Lang$^{1,2}$, A. D. Martin$^{3}$, R. S. Thorne$^{2}$, \\                                                 
                                                   
\vspace*{0.5cm}
${}^1${\it Rudolf Peierls Centre for Theoretical Physics, University of Oxford,\\  1 Keble Road, OX1 3NP, UK}\\
${}^2${\it Department of Physics and Astronomy, University College London, WC1E 6BT, UK}   \\                                                    
${}^3${\it Institute for Particle Physics Phenomenology, University of Durham, DH1 3LE, UK} \\                                              
                                                    
\vspace*{1cm}                         

\begin{abstract}

\noindent  We investigate the impact of the high precision ATLAS and CMS 7 TeV measurements of inclusive jet production on the MMHT global PDF analysis at next--to--next--to--leading order (NNLO). This is made possible by the recent completion of the long--term project to calculate the NNLO corrections to the hard cross section. We find that a  good description of the ATLAS data is not possible with the default treatment of experimental systematic errors, and propose a simplified solution that retains the dominant physical information of the data. We then investigate the fit quality and the impact on the gluon PDF central value and uncertainty when the ATLAS and CMS data are included in a MMHT fit. We consider  both common choices for the factorization and renormalization scale, namely the inclusive jet transverse momentum, $p_\perp$, and the leading jet $p_\perp$, as well as the different jet radii for which the ATLAS and CMS data are made available. We find that the impact of these data on the gluon is relatively insensitive to these inputs, in particular the scale choice, while the inclusion of NNLO corrections tends to improve the data description somewhat, and gives a qualitatively similar though not identical impact on the gluon in comparison to NLO.

\end{abstract}
                                   
\end{center}

\section{Introduction}

Parton distribution functions (PDFs) are a fundamental input into hadron collider physics for both the theoretical and experimental particle physics communities, see~\cite{Gao:2017yyd} for a recent review. The dominant experimental input in determining PDFs comes from data on deep inelastic scattering (DIS) structure functions, with the final combined HERA Run I+II dataset being the most prominent example~\cite{Abramowicz:2015mha}. By combining data from different processes (neutral and charged current) and targets (protons, deuterons, heavy nuclei) one can obtain much direct information about the quark content of the proton, while the quark evolution is sensitive to the gluon. Indeed, at small $x$ this evolution is largely driven by the gluon. However, at moderate and high $x$ the proton structure is dominated by the non--singlet valence quark distributions. Then, the evolution is largely decoupled from the gluon and the speed of evolution is determined mostly by the value of the strong coupling constant. There is of course some influence from the gluon in the high--$x$ quark evolution; it provides an increase of quarks (and anti--quarks) with $Q^2$, but this is difficult to decorrelate from the variation with $\alpha_S$, and also decreases in significance at high $x$. 

In order to obtain the most comprehensive constraints various groups perform global fits to all available data for which precise theoretical calculations are available~\cite{Harland-Lang:2014zoa,Dulat:2015mca,Ball:2017nwa}. For example, Drell Yan data in hadron--hadron collisions (both collider and fixed target) provide additional information on the anti--quarks and on the quark flavour decomposition, and improve the overall determination in comparison to DIS data alone~\cite{Accardi:2016qay,Alekhin:2017kpj}. A direct constraint on the gluon, particularly at high $x$, can be obtained from high $p_\perp$ jet production data at hadron colliders\footnote{Constraints from differential top quark pair production~\cite{Czakon:2016olj} and the $Z$ boson $p_\perp$ distribution~\cite{Boughezal:2017nla} have recently become feasible, in particular due to the relatively recent release of high precision LHC data and NNLO theory calculations for these processes.}. However, until recently the data have been quite limited in precision, while the calculation of the hard cross section has only been available up to next-to-leading order (NLO), with some threshold resummation results also available~\cite{Kidonakis:2000gi,Kumar:2013hia,deFlorian:2013qia,Liu:2017pbb}. In our most recent global fit~\cite{Harland-Lang:2014zoa} we included data on inclusive jet production as a function of $p_T$ in different rapidity bins from the D0~\cite{Abazov:2011vi} and CDF~\cite{Abulencia:2007ez} experiments at the Tevatron, and from early measurements at both the ATLAS~\cite{Aad:2011fc,Aad:2013lpa} and CMS~\cite{Chatrchyan:2012bja} detectors, at 7~TeV. The Tevatron data were generally close to threshold, so that we could reliably include the NNLO approximations obtained from expanding out the threshold resummation of~\cite{Kidonakis:2000gi}. However, the LHC data extended much further from threshold, where it was clear these approximations break down~\cite{deFlorian:2013qia}. Thus in that study we only included the Tevatron jet data in the NNLO fit. We note that at this time other groups used alternative approaches of including jet data at NNLO, see~\cite{Ball:2014uwa,Dulat:2015mca}. 
  
Since this previous study there has been both an increase in the range and precision of LHC jet data, combined with the completion of a very large--scale and long--term project to calculate the NNLO corrections to the hard cross section~\cite{Ridder:2013mf,Currie:2013dwa,Currie:2016bfm}. In this paper we investigate the consequences of both of these new developments for PDF determination, concentrating on the final 7 TeV measurements from the ATLAS~\cite{Aad:2014vwa} and CMS~\cite{Chatrchyan:2014gia} collaborations. We find that neither is quite as straightforward as might be hoped. First, the newer ATLAS jet data are impossible to fit well without some modifications. Second, the variation in NNLO corrections between scale choices and jet radii is potentially quite significant. We therefore examine both of these issues in detail, considering the impact of different choices of scale and jet radius on the fit at both NLO and NNLO, while suggesting a minimal manner in which to improve the fit quality in the case of the ATLAS data, that retains the dominant physical constraints implied by the data. We then determine the consequences for both the central values and uncertainties of the gluon PDF obtained at both NLO and NNLO within the MMHT framework. We obtain the very encouraging, and not necessarily expected, result that in practice these are found to very insensitive to any reasonable choices we make in either the treatment of the data or the theory input\footnote{As such, a detailed investigation of the scale dependence as described in~\cite{Martin:2017uyr} may be avoided.}. We also find in general that the data description is somewhat improved by the inclusion of these NNLO corrections. As mentioned above, we only consider the 7 TeV data, for which the NNLO calculations are currently available, in this study. In fact, already a range of precise jet data from ATLAS and CMS at 8 and 13 TeV~\cite{Aaboud:2017dvo,Aaboud:2017wsi,Khachatryan:2016mlc,Khachatryan:2016wdh} are available. This study will therefore guide the inclusion of these data in future MMHT fits at NNLO. For example, a similarly poor default description is also present in the ATLAS 8 TeV~\cite{Aaboud:2017dvo} and 13 TeV~\cite{Aaboud:2017wsi} data, and so must be dealt with in any future fit.

The outline of this paper is as follows. In Section~\ref{sec:theory} we describe the theoretical calculation and tools used. In Section~\ref{sec:ATLAS} we describe the issues related to  the fit to the ATLAS data and develop a simplified approach to improve the data description. In Section~\ref{sec:NNLOfit} we study the fit quality at NLO and NNLO to the ATLAS and CMS 7 TeV jet data, for different choices of jet scale and radius. In Section~\ref{sec:PDF} we show the impact of these data on the central value and uncertainty of the gluon PDF. Finally, in Section~\ref{sec:conc} we conclude.
  
\section{Theoretical inputs for jet production}\label{sec:theory}

\begin{figure}
\begin{center}
\includegraphics[scale=0.66]{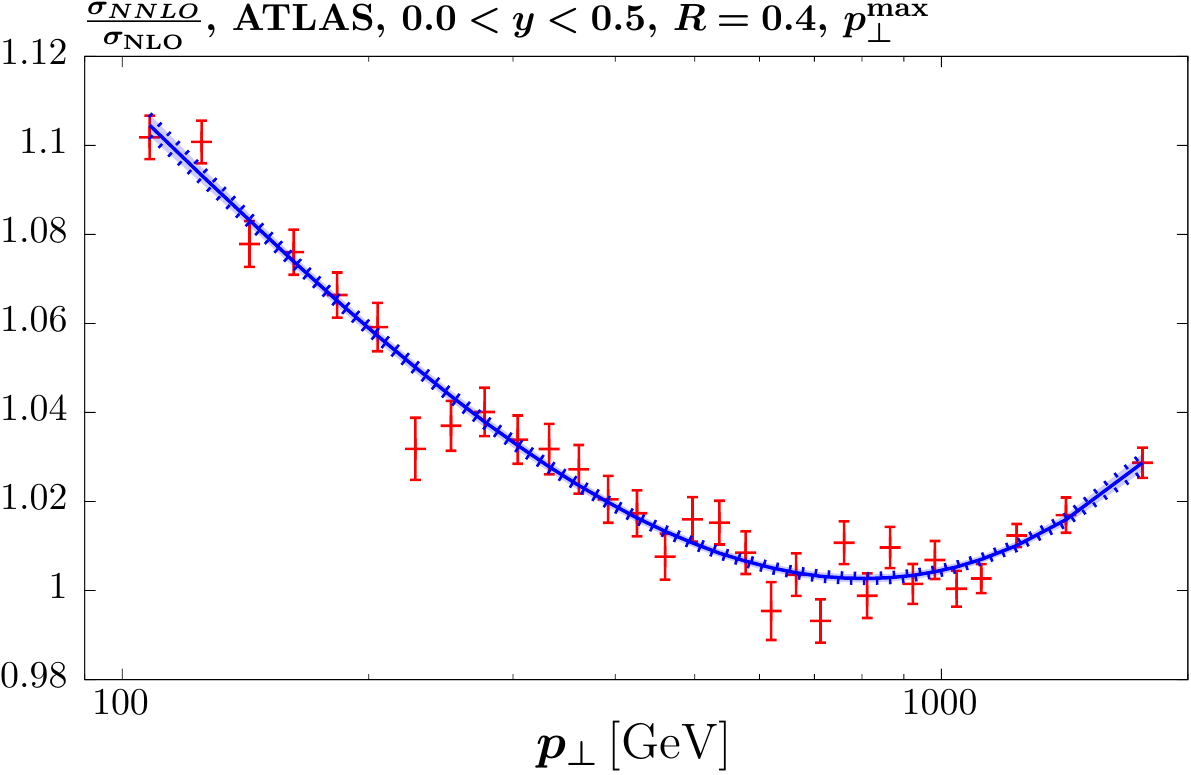}
\includegraphics[scale=0.66]{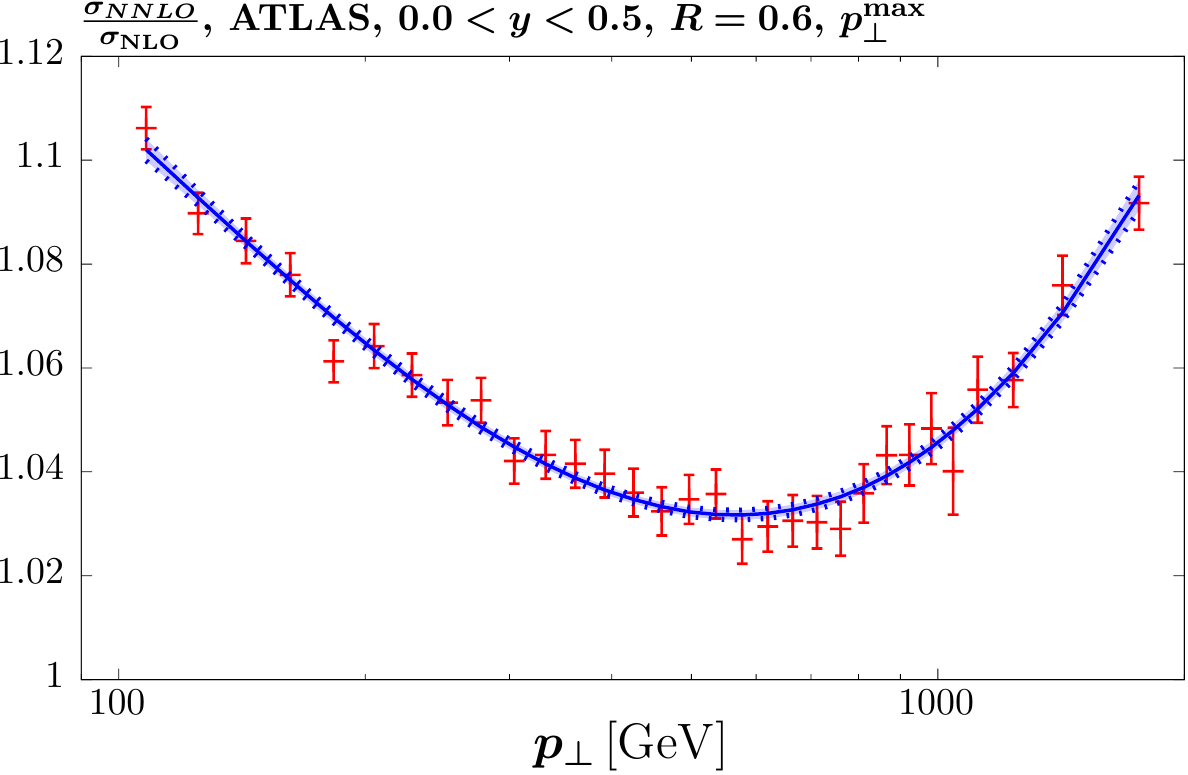}
\includegraphics[scale=0.66]{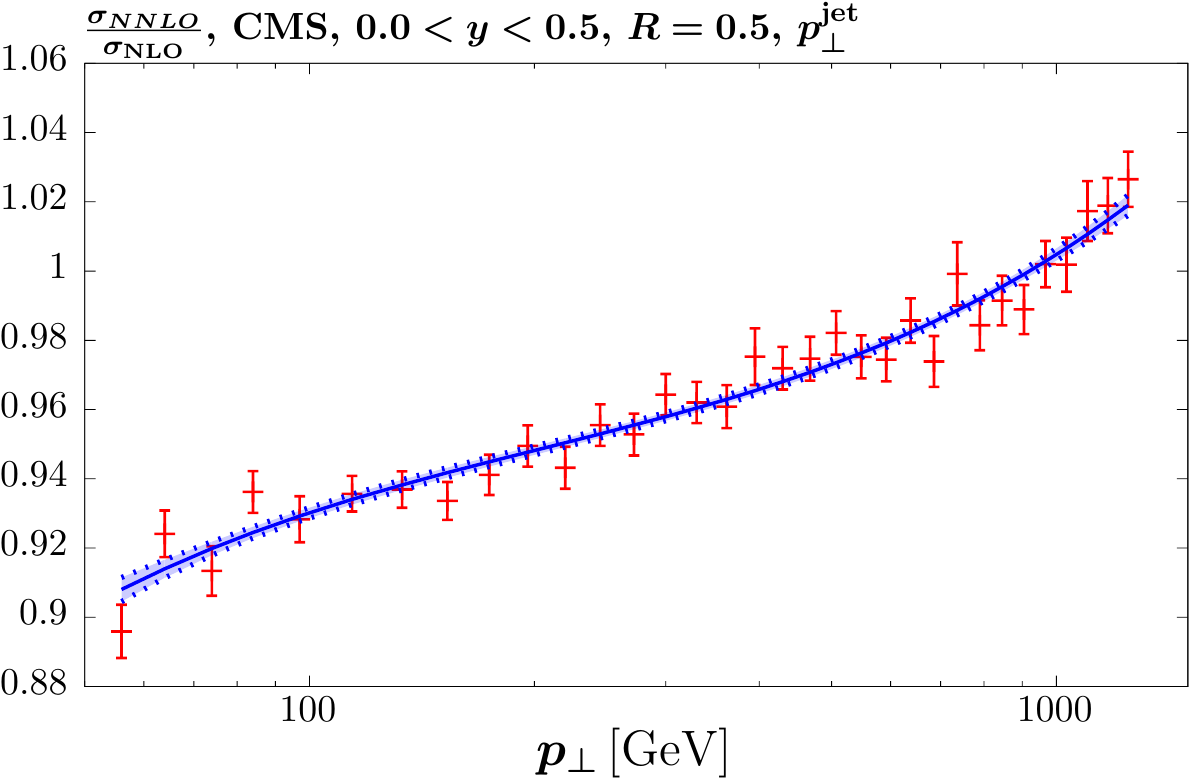}
\includegraphics[scale=0.66]{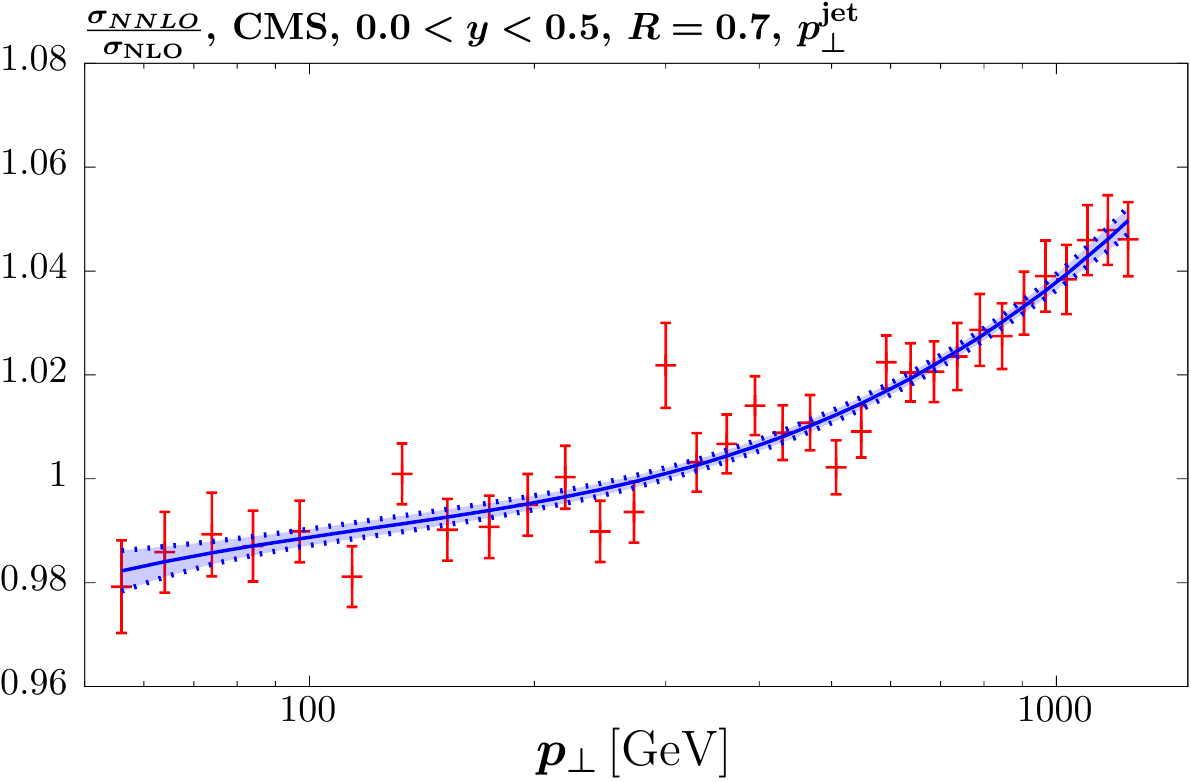}
\caption{The calculated NNLO to NLO K-factors~\cite{Currie:2016bfm}, including the MC statistical errors, for representative ATLAS and CMS 7 TeV jet kinematics. Also shown is the 4--parameter fit to the K-factors described in the text. The uncertainty band is shown for illustration by summing in quadrature the 68\% C.L. fit uncertainties in each bin, i.e. omitting correlations. The top left (right) plots show the ATLAS result with the $p_\perp^{\rm max}$ scale choice and $R=0.4$ (0.6), while the bottom left (right) plots show the CMS result with the $p_\perp^{\rm jet}$ scale choice and $R=0.5$ (0.7). In both cases results for the central rapidity bin are shown.}
\label{fig:kfacs}
\end{center}
\end{figure} 

The NLO theoretical predictions for inclusive jet production are calculated using the \texttt{NLOJet++} code~\cite{Nagy:2003tz}, with the results stored in \texttt{APPLgrid}~\cite{Carli:2010rw} format\footnote{In the case of the ATLAS data with the $p_\perp^{\rm max}$ scale we take this directly from the \texttt{APPLgrid} website~\cite{Applgridweb}.} for fast use in the PDF fit. The theoretical approach used to calculate the NNLO corrections to this is described in~\cite{Currie:2016bfm} (see also~\cite{Currie:2017ctp}). The NNLO to NLO K-factors are provided by the authors for the ATLAS and CMS 7 TeV kinematics, for both jet radii presented by the collaborations and with the renormalization/factorization scale taken as either inclusive jet transverse momentum, $p_\perp^{\rm jet}$, or the maximum jet transverse momentum, $p_\perp^{\rm max}$. These are provided using the NNPDF3.0 set~\cite{Ball:2017nwa}, although the K--factors are expected to be largely insensitive to the PDF choice. This therefore allows us to perform a detailed analysis of the impact of these LHC jet data for different jet radii and scale choices. 

The predictions are provided with a statistical uncertainty due to the MC integration in the theoretical calculation. We show the predictions for the central ATLAS and CMS rapidity bins for illustration in Fig.~\ref{fig:kfacs}, taking representative choices of the jet radius and scale. We can see that the MC uncertainties are up to 1\% of the K--factors, while the central values are at most 10\% from unity. These errors are therefore non-negligible and must be included in the fit. One possibility is to simply include these as an additional bin--by--bin source of uncorrelated uncertainty. However in general we will expect the K--factors to be smoothly varying functions of the kinematic variables, and therefore this approach is unnecessarily conservative. We instead perform a simple 4--parameter fit 
\be
\frac{\sigma_{\rm NNLO}^i}{\sigma_{\rm NLO}^i}=\lambda_0^{i}+\lambda_1^i \log p_\perp+\lambda_2^i \log^2 p_\perp+\lambda_3^i \log^3 p_\perp\;,
\ee
where the `$i$' labels the specific rapidity region, data set, choice of jet radius and scale. In general this enables a good fit, with $\chi^2/{\rm dof}\sim 1$, and the standard  $\Delta \chi^2=1$ criterion can be applied to determine the uncertainties associated with the fit. This results in four sources of correlated systematic uncertainty for each rapidity bin, which we then include in the PDF fit. We have explicitly checked that the results are stable with respect to adding in further polynomial terms, and thus we can safely truncate at this order. The best fit curves are shown in Fig.~\ref{fig:kfacs} for four representative cases, with the uncertainty band due to the sum in quadrature of the errors in each bin, i.e. omitting correlations, shown for illustration. We can see that biggest difference in the K--factors is due to the choice of scale, which gives quite different trends depending on whether $p_\perp^{\rm max}$ or $p_\perp^{\rm jet}$ is taken. The choice of jet radius also shows some impact, while the ATLAS and CMS results with the same scale choice and comparable (either low or high) jet radii,  which are not shown here, show a qualitatively similar trend.

\section{Treatment of ATLAS correlated systematic errors}\label{sec:ATLAS}

Before considering the general impact of the jet data on the NNLO fit, some care is needed when dealing with the ATLAS data~\cite{Aad:2014vwa}. The discussion follows closely that given in~\cite{Harland-Lang:2017dzr}, but for completeness we present a summary below. For illustration, we will work at NLO only, using as a baseline PDF the MMHT14 set~\cite{Harland-Lang:2014zoa} including the HERA I+II combined data~\cite{Abramowicz:2015mha}, that is as presented in~\cite{Harland-Lang:2016yfn}. 

\begin{figure}
\includegraphics[scale=0.55]{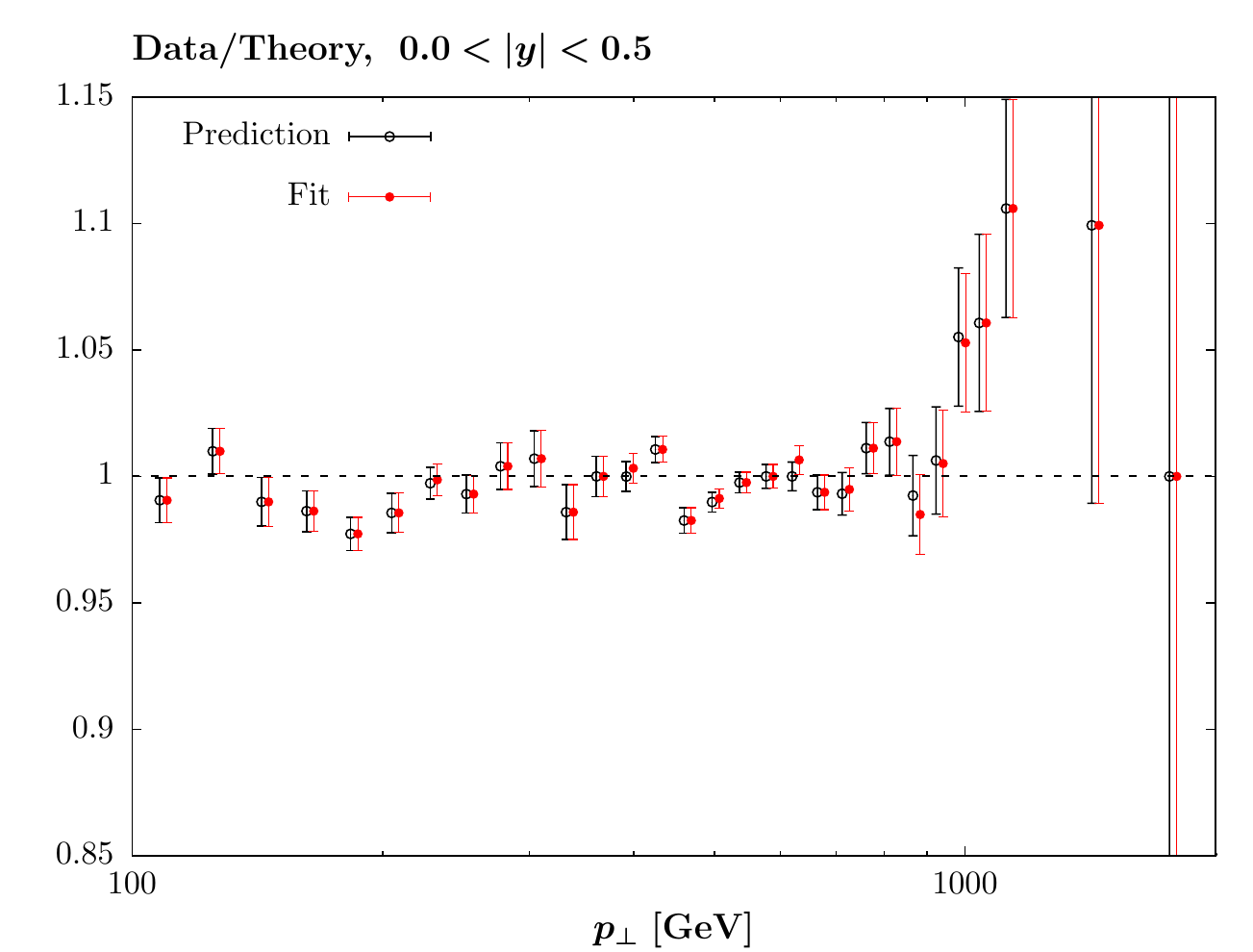}\qquad
\includegraphics[scale=0.53]{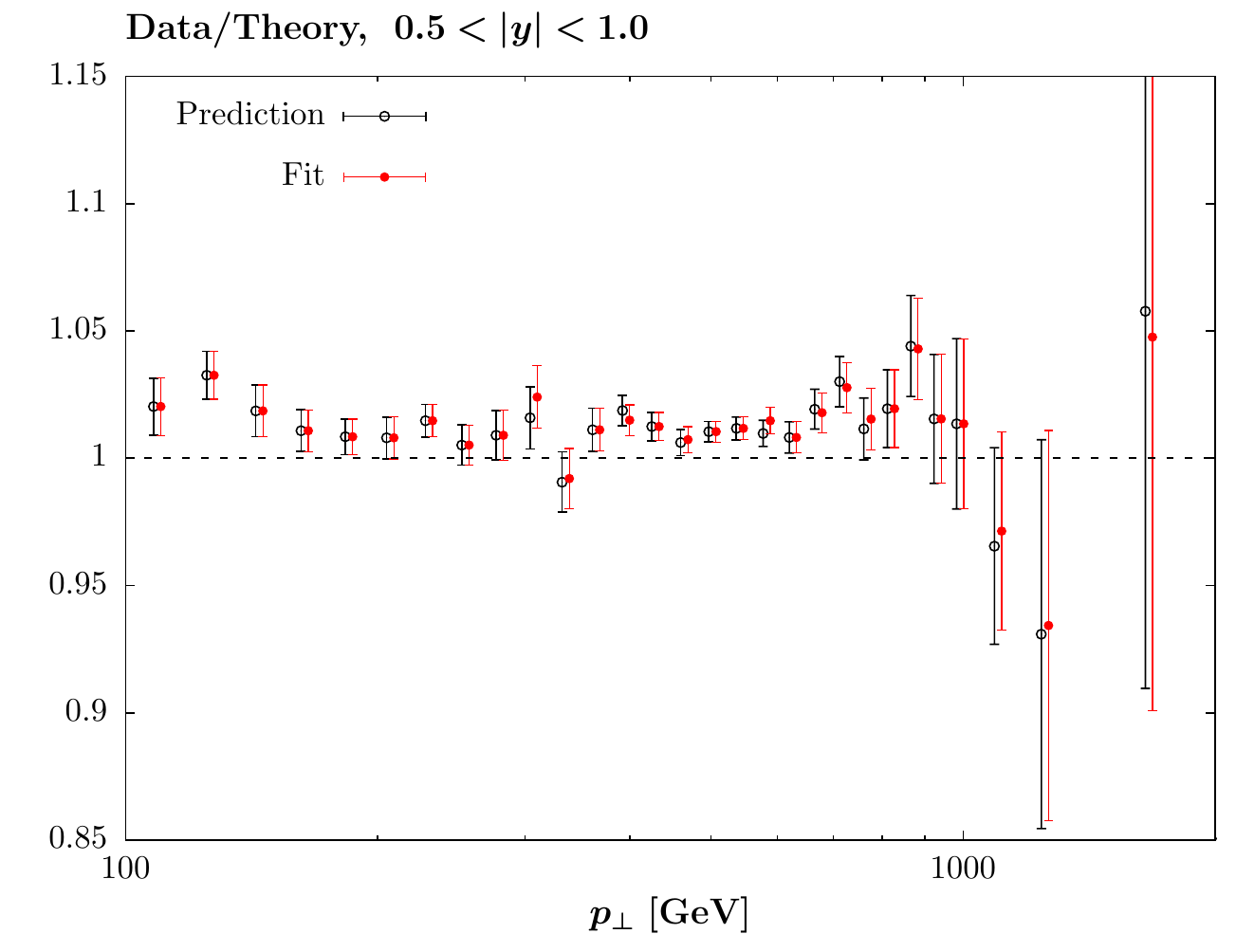}
\includegraphics[scale=0.55]{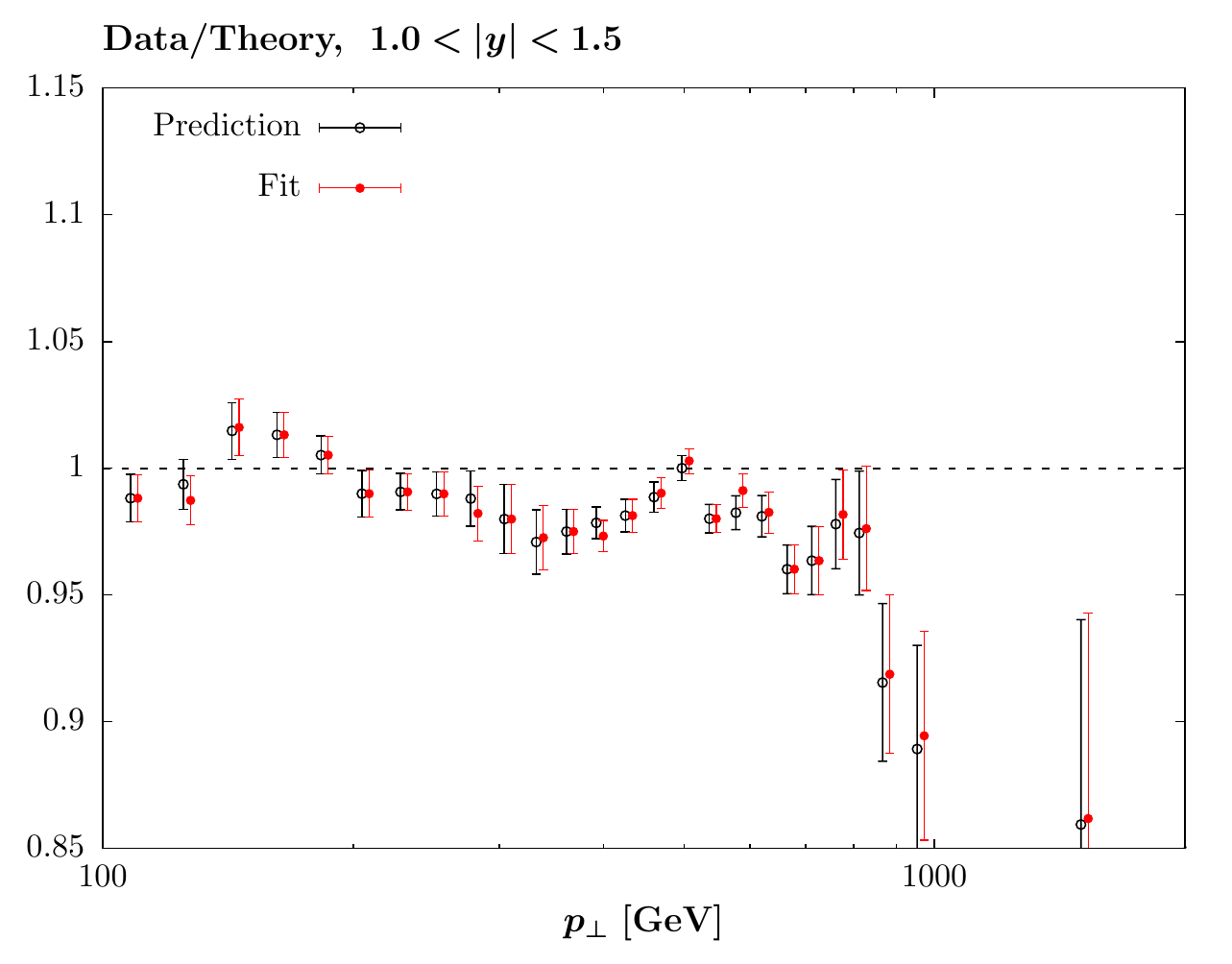}\qquad
\includegraphics[scale=0.55]{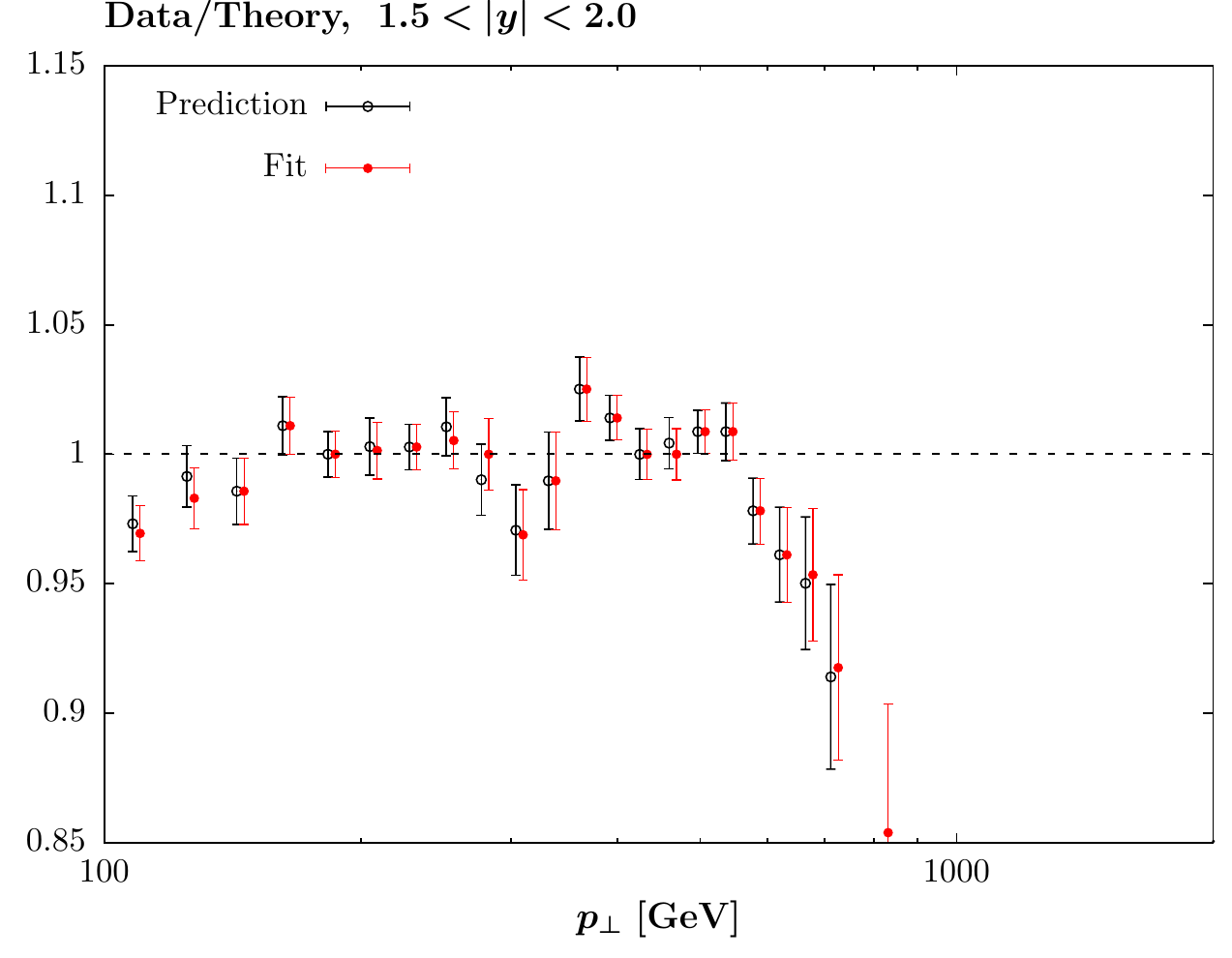}
\caption{Comparison of NLO prediction and fit to ATLAS jet data~\cite{Aad:2014vwa} for first four rapidity bins. Data/theory is plotted, with the data already shifted by the systematic uncertainties in order to achieve the best description. The displayed errors are purely statistical.}
\label{fig:jet1}
\end{figure}

The predicted and fit data/theory for the first four jet rapidity bins are shown in Fig.~\ref{fig:jet1}, with the shifts due to the correlated systematic uncertainties included and 
using $R=0.4$ and $p_\perp^{\rm jet}$. The description of the data is visibly poor, in particular in the $0.5 < |y_j|<1.0$ and $1.0<|y_j|<1.5$ bins across all $p_\perp$, and with some deterioration in the other bins at high $p_\perp$.  A significant contributing factor to this is an essentially systematic offset in the data/theory between these neighbouring rapidity bins, but in opposite directions; as these probe PDF sets of the same flavour in very similar $x$ and $Q^2$ regions little improvement is possible (or observed) by refitting to these data. 

\begin{figure}
\centerline{%
\includegraphics[scale=0.55]{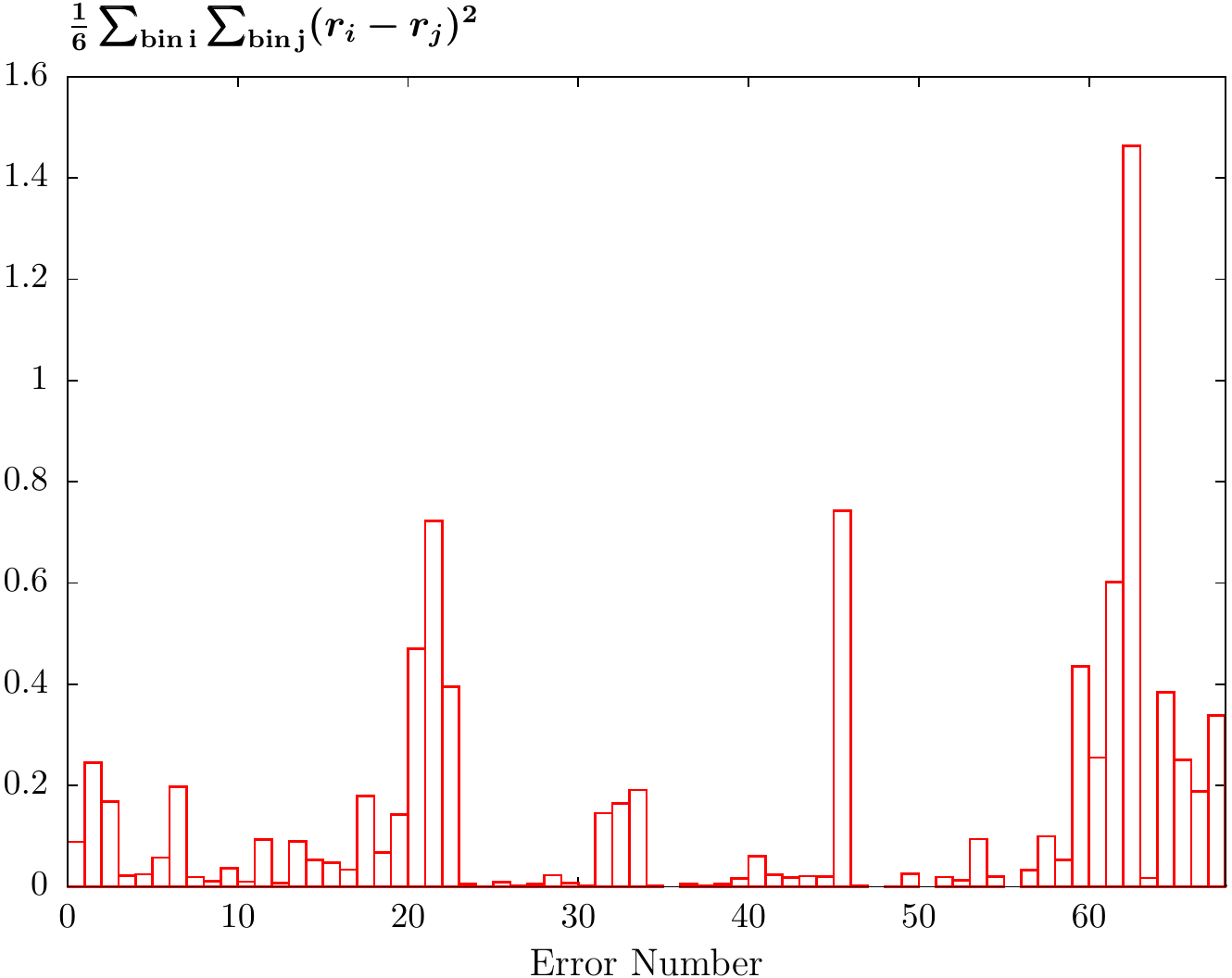}}
\caption{Average squared sum of the systematic shift differences $(r_i-r_j)^2$ for the first four rapidity bins of the ATLAS 7 TeV jet data~\cite{Aad:2014vwa}.}
\label{fig:jetshifts}
\end{figure}

The cause of this appears to lie with the shift allowed by the correlated systematic uncertainties. The ATLAS data contain a large number of individual correlated errors which are generally completely dominant over the (small) statistical errors; for the `weaker' assumption about error correlations defined in~\cite{Aad:2014vwa} that we take, there are 71 individual sources of systematic error. If we simply assume that all of these uncertainties are completely decorrelated between the six rapidity bins (while remaining fully correlated within the bins) a universally good description is found: in this case, the extra freedom allows the data to shift in order to achieve a reasonable data/theory description. This is however clearly a hugely over--conservative assumption. To be more precise, we examine the size of the shifts $r_k$ for each source of systematic uncertainty by which the theory (or equivalently, data) points are allowed to move, as defined in the $\chi^2$
\begin{equation}\label{eq:chicor}
\chi^2=\sum_{i=1}^{N_{\rm pts}}\left(\frac{D_i+\sum_{k=1}^{N_{\rm corr}}r_k\sigma_{k,i}^{\rm corr}-T_i}{\sigma_i^{\rm uncorr}}\right)+\sum_{k=1}^{N_{\rm corr}} r_k^2\;,
\end{equation}
where $D_i$ is $i$th data point, $T_i$ is the theory prediction and $\sigma^{\rm uncorr}_i$ ($\sigma^{\rm corr}_{k,i}$) are the uncorrelated (correlated) errors.
We in particular evaluate the shifts for each of the first four rapidity bins (from 0 to 2.0 in steps of 0.5) individually; including the last two rapidity bins, where the data tend to be less precise, does not affect the conclusions that follow. Any tensions between the different bins may then show up through significantly different $r_k$ values being preferred in the different rapidity bins, in order to achieve good individual fits. In Fig.~\ref{fig:jetshifts} we show the average squared sum of the shift differences $(r_i-r_j)^2$ for the four bins. It is clear that for a small subset  of the shifts the size of this difference is significantly larger than zero, indicating a large degree of tension. 

The three shifts \texttt{jes21}, \texttt{45} and \texttt{62} as defined in~\cite{Aad:2014bia}, which correspond~\cite{ATLASpriv} to the multi-jet balance asymmetry, an in--situ statistical uncertainty and the jet energy scale close by jets, respectively, show particularly large differences. However, for the in--situ statistical uncertainty the correlations are particularly well determined~\cite{ATLASpriv}, and therefore we omit this from the investigation below. In fact, as shown in~\cite{Harland-Lang:2017dzr}, decorrelating this source of uncertainty has a less significant impact on the quality of the data description in comparison to the other two sources.

\begin{figure}
\centerline{%
\includegraphics[scale=0.5]{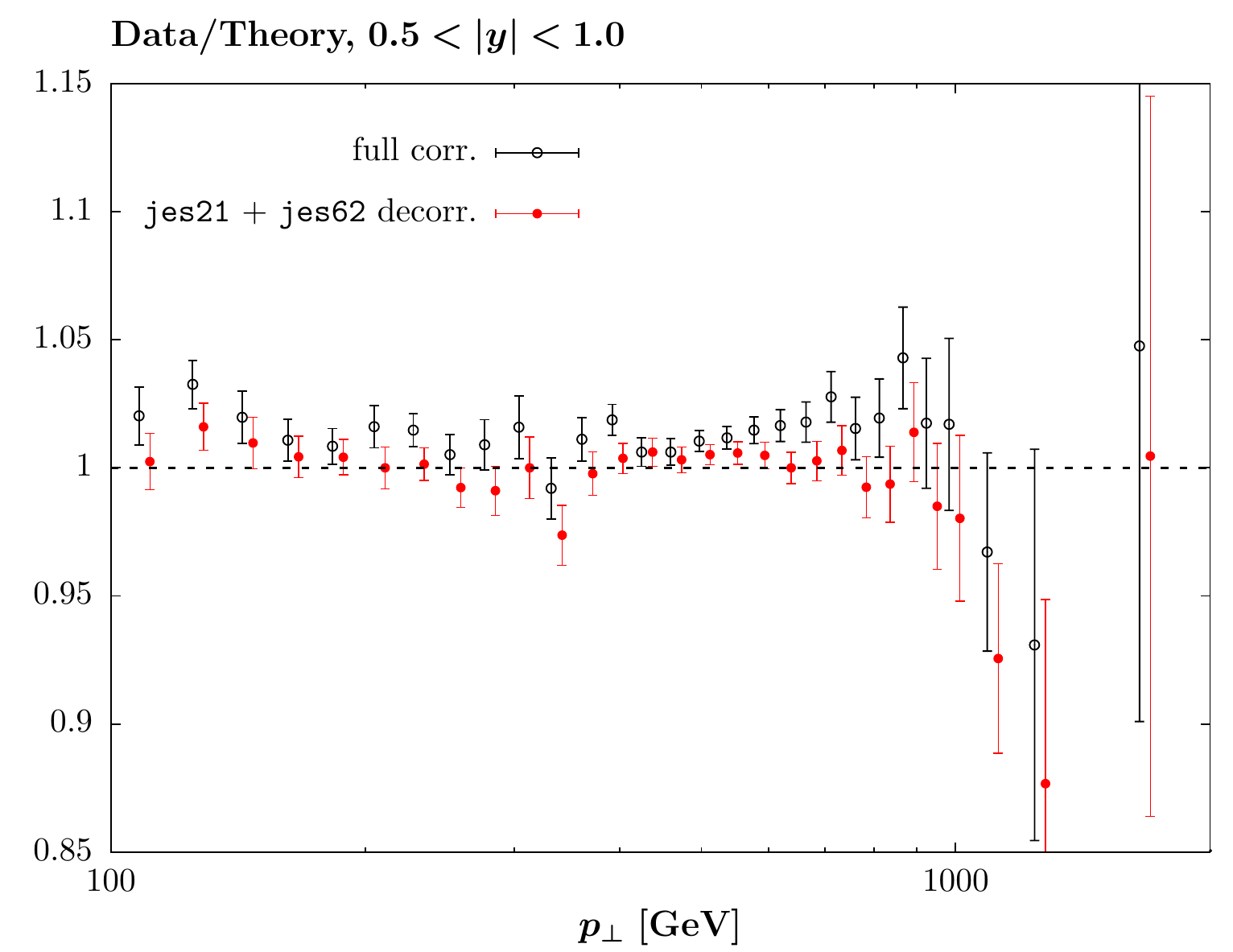}\qquad
\includegraphics[scale=0.5]{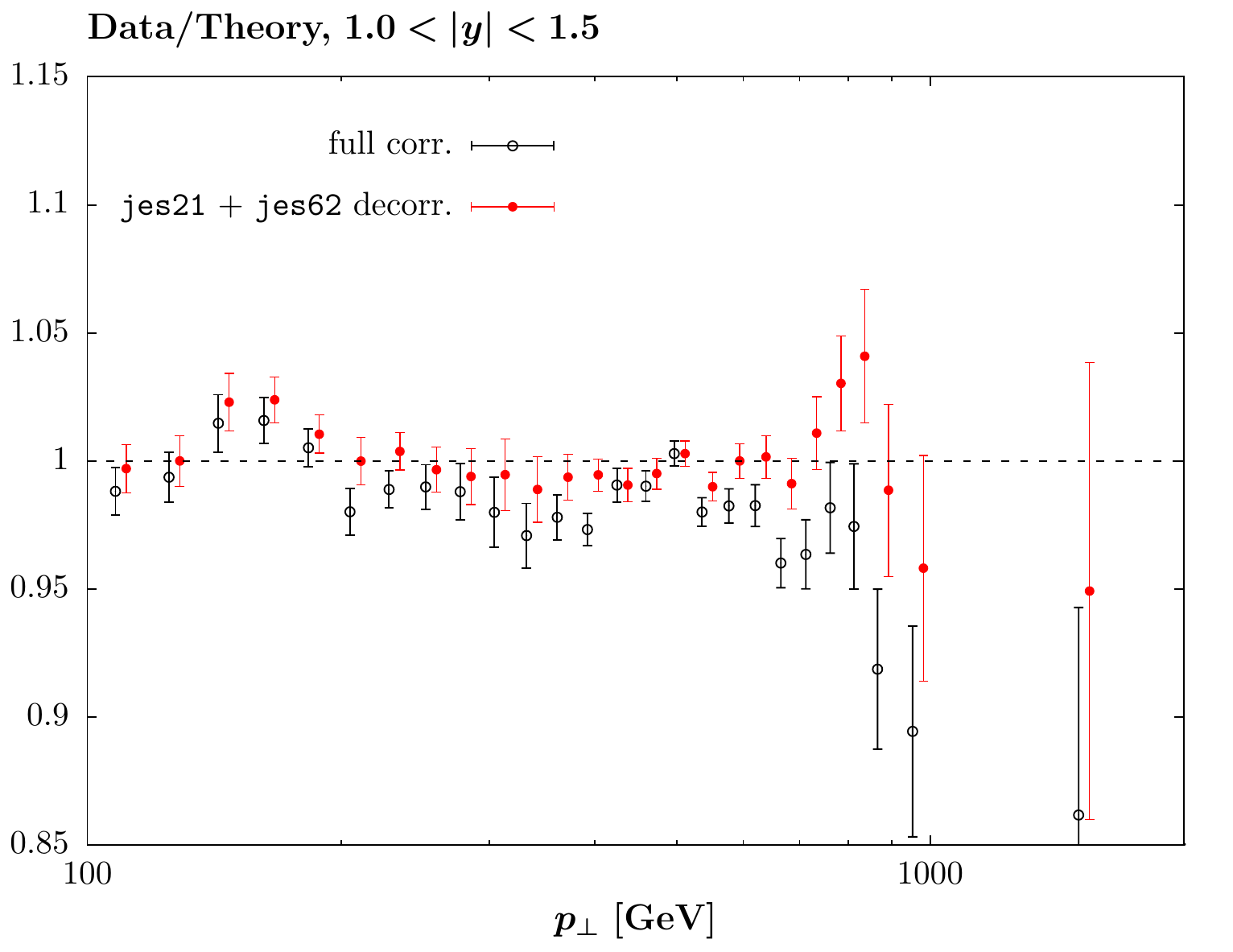}}
\caption{Data/theory fit as in Fig.~\ref{fig:jet1}, for $0.5<|y_j|<1.0$ and $1.0<|y_j|<1.5$, with and without the labelled systematic errors decorrelated between jet rapidity bins.}
\label{fig:jet2}
\end{figure}

We therefore investigate the impact of decorrelating the systematic uncertainties, \texttt{jes21} and \texttt{62}, alone between rapidity bins. We compare to the ATLAS data with $R=0.4$ and $p_\perp^{\rm jet}$ as the scale choice. The result for the individual uncertainty sources, as well as the combination, is shown in Table~\ref{tab:decor},  and is found to be dramatic. Simply decorrelating \texttt{jes21}, for example, leads to a reduction of $180$ points in $\chi^2$, giving almost a factor of 2 decrease in the $\chi^2/N_{\rm pts.}$ from 2.85 to 1.58. Decorrelating \texttt{jes62} in addition gives a $\chi^2/N_{\rm pts.}$ of 1.27. The same data/theory comparisons as in Fig.~\ref{fig:jet1}, but including this decorrelation of \texttt{jes21} and \texttt{jes62}, are shown in Fig.~\ref{fig:jet2} and are visibly improved, with the additional freedom allowing the data/theory to shift in the different rapidity bins and achieve a good overall description. While the above analysis only considers the experimental sources of correlated uncertainty, we have also checked that decorrelating the quoted uncertainty associated with the non--perturbative corrections from~\cite{Aad:2014vwa} that we apply leads to some $\sim 40$ point improvement in the $\chi^2$, that is significantly smaller than for those sources discussed above. In addition, we find that even omitting these corrections entirely has little impact on the fit quality, in other words these appear to be correlated sufficiently with other sources of experimental systematics that their omission does not affect the comparison significantly

We note that this corresponds to a simplified version of the alternative correlation scenarios presented in~\cite{Aaboud:2017dvo} subsequently to the discussion in~\cite{Harland-Lang:2017dzr}. Here, the impact of a more conservative partial decorrelation of various sources of uncertainty (including theoretical uncertainties due to scale choice and variation) in the 8 TeV ATLAS jet data is investigated, and a comparable although somewhat less dramatic improvement in the data description quality is found. However, as we will show below, the effect of our simplified decorrelation model is to improve the fit quality while having a limited effect on the PDFs themselves. Therefore we do not expect the details of the decorrelation model to have a significant impact on the final result. Thus, while the correlation between systematic errors should clearly be determined by physics considerations and not simply the possibility of improving the theory description of the data, the simplified approach we take is sufficient for our purposes.

\begin{table}
\begin{center}
\begin{tabular}{|c|c|c|c|c|c|c|}
\hline
&Full&\texttt{21}&\texttt{62}&\texttt{21,62}\\
\hline
$\chi^2/N_{\rm pts.}$&2.85&1.58&2.36&1.27\\
\hline
\end{tabular}
\caption{$\chi^2$ per number of data points ($N_{\rm pts}=140$) for fit to ATLAS jets data~\cite{Aad:2014vwa}, with the default systematic error treatment (`full') and with certain errors, defined in the text, decorrelated between jet rapidity bins.}\label{tab:decor}
\end{center}
\end{table}

\section{Fit quality at NNLO}\label{sec:NNLOfit}

\subsection{Individual data sets at NNLO}

In Table~\ref{tab:ATLASnnlo} we show the quality, $\chi^2$, of the prediction and fit to the ATLAS and CMS jet data. For the predictions, we take as a baseline set the fits to the same data set (and using the same theoretical parameters) as MMHT14~\cite{Harland-Lang:2014zoa}, but including the final HERA I+II combined data set~\cite{Abramowicz:2015mha}, and excluding all Tevatron jet data. In the latter case the NNLO predictions are not currently publicly available and so these are omitted for consistency. Unless otherwise stated we take $p_\perp^{\rm jet}$ as the factorization/renormalization scale. The NLO (NNLO) results are all made with a fixed value of $\alpha_s$ of 0.120 (0.118), as taken in~\cite{Harland-Lang:2014zoa}, although the results are insensitive to this precise choice. We first consider the impact of fitting the ATLAS and CMS jet data individually. We show the `ATLAS' result with the default treatment of systematic errors, with our model of partial error decorrelation ($\sigma_{pd}$), and with a full decorrelation of all systematic errors across jet rapidity bins ($\sigma_{fd}$). While as discussed above this latter approach is clearly overly conservative, we note that e.g. only fitting the first jet rapidity bin as in~\cite{Ball:2017nwa} implicitly assumes such a decorrelation.

As in the NLO case above, the description and fit of the ATLAS data with the default error treatment is poor, with $\chi^2/N_{\rm pts}\sim 2$ or higher, but this improves to be of order unity when taking our model of partial error decorrelation. If the systematic errors are fully decorrelated between rapidity bins, some further improvement is achieved, giving a value that is somewhat below unity. However, it is clear that the most dramatic change comes from the decorrelation of the first two systematic errors. We also show the comparison  for different choices of jet radius, with $R=0.4 (0.5)$ and $R=0.6 (0.7)$ for the ATLAS(CMS) data, which in the following we will label as `low' and `high', respectively. Interestingly, with the higher choice of $R$ the quality of the description of the ATLAS data is better, while the change when refitting is significantly increased; for the partial error decorrelation the $\chi^2$ decreases by $\sim 30$ points, giving a final $\chi^2/N_{\rm pts}$ very close to unity. On the other hand, for the full error decorrelation, little difference is seen, which is perhaps unsurprising given the over--estimate in the freedom of the data uncertainties. We also show the $\chi^2$ for the prediction and fit to the CMS jet data. Here the description is fair, and a  $\chi^2/ N_{\rm pts}\sim 1$ is achieved for both radii after refitting, with a reduction in  the $\chi^2$ by $\sim 30$ points. The fit quality is a little better for the lower choice of jet radius, although the difference is relatively small.

\begin{table}
\begin{center}
\renewcommand\arraystretch{1.15}
\begin{tabular}{|c|c|c|c||c|c|c|}
\hline
& ATLAS & ATLAS, $\sigma_{pd}$ & ATLAS, $\sigma_{fd}$&& CMS \\
\hline 
$R=0.4$& 350.8 (333.7)  & 183.1 (170.7)  & 128.4 (122.2) &$R=0.5$ &191.7 (163.4) \\
\hline 
$R=0.6$& 304.0 (264.0) & 178.8 (148.9) & 128.9 (115.7) & $R=0.7$&  200.1 (175.2)  \\
\hline
\end{tabular}
\caption{The $\chi^2$ for the ATLAS ($N_{\rm pts}=140$) and CMS 7 TeV jet data ($N_{\rm pts}=158$) at NNLO. The quality of the description using the baseline set is shown, while the result of re--fitting to the single jet data set is given in brackets. Results with the different treatments of the ATLAS systematic uncertainties, described in the text, are also shown.}
\label{tab:ATLASnnlo}
\end{center}
\end{table}

\subsection{NNLO vs. NLO fit quality for combined data}
 
In Table~\ref{tab:combnnlo} we show the results of the combined fit to the ATLAS and CMS data, taking the partial error decorrelation model for the ATLAS data. As above, we show results for low and high jet radii, while we also consider the impact of the jet scale choice. The $\chi^2$ values for the ATLAS and CMS datasets are given.

\begin{table}
\begin{center}
\renewcommand\arraystretch{1.15}
\begin{tabular}{|c|c|c|c|c|}
\hline
& $R_{\rm low}$, $p_\perp^{\rm jet}$ &$R_{\rm low}$, $p_\perp^{\rm max}$& $R_{\rm high}$, $p_\perp^{\rm jet}$&$R_{\rm high}$, $p_\perp^{\rm max}$\\
\hline 
ATLAS (NLO)& 213.8 & 190.5 & 171.5 & 161.2\\
\hline
ATLAS (NNLO) & 172.3  & 199.3 &149.8 & 152.5 \\
\hline 
CMS (NLO)& 190.3 & 185.3& 195.6& 193.3\\
\hline
CMS (NNLO) & 177.8 & 187.0 & 182.3 & 185.4  \\
\hline
\end{tabular}
\caption{The $\chi^2$ for the combined fit to the ATLAS ($N_{\rm pts}=140$) and CMS ($N_{\rm pts}=158$) 7 TeV jet data. The values for the ATLAS and CMS contributions are given, for different choices of jet radius and scale, at NLO and NNLO.}
\label{tab:combnnlo}
\end{center}
\end{table}

A number of observations can be made. First, the quality of the description when fitting both data sets simultaneously is nearly as good as when fitting each individually. This is particularly true for ATLAS, while for CMS there is a little more deterioration. Thus, there is no significant tension between the ATLAS and CMS jet data. The deterioration in the fit quality for the other datasets (not shown here), is relatively mild, with $\Delta\chi^2\sim $ 5 (10) for the low (high) jet radius choice. The biggest deterioration is in the BCDMS deuteron data for the high jet radius choice, which deteriorates by $\sim 7$ units. 

Second, at NNLO some improvement in the fit quality is apparent for most choices of jet radii and scale; such an improvement is also visible upon comparing Tables~\ref{tab:decor} and~\ref{tab:ATLASnnlo} for the individual fits. This is more significant for the ATLAS data, where an improvement of up to 40 points in $\chi^2$ can be achieved, while for the CMS data the improvement is at most 10 points. On the other hand, for the low jet radius, and $p_\perp^{\rm max}$ choice, some slight deterioration in the fit quality is observed. We note that in~\cite{Harland-Lang:2017dzr} a deterioration in the fit quality to the ATLAS data when going to NNLO was reported, however here it was precisely these choices of scale and jet radius that were taken. Following the more detailed study in this work, we can see that this effect is not in general present. 

Third, we can see that for the joint fit a clear preference for the higher choice of jet radius is shown at both orders in the ATLAS data, while for the CMS any difference is relatively marginal. Moreover, while at NLO some preference (in particular in the ATLAS data) for the $p_\perp^{\rm max}$ scale choice is shown, at NNLO this trend is reversed for the low $R$ choice, while for high $R$ essentially no preference is indicated by the fit, with the descriptions of the ATLAS and CMS data being excellent for both scale choices. Thus to achieve the best NNLO fits to these data sets, a higher value of $R$ is preferred, while the result is less sensitive to the choice of scale. As we will show in the following section, this relative insensitivity is also observed in the extracted PDFs, in particular for the gluon.

Finally, it is important to clarify the role played by the NNLO jet production theory, in contrast to the NNLO PDFs, in leading to the improvement in the fit quality at NNLO. In Table~\ref{tab:combnnlonk} we show the same $\chi^2$ values as before, resulting from the NNLO fit to the combined ATLAS and CMS data, but in addition excluding the NNLO K--factors, i.e. applying NLO theory only to the jet data. We can see that the improvement due to the NNLO corrections in the fit is still present at roughly the same level as before, with some variation in the precise amount. We also show the effect of excluding the correlated errors associated with the K--factor fit described in Section~\ref{sec:theory}. This leads to some small increase in the $\chi^2$, as it must, but the trend is unchanged.

\begin{table}
\begin{center}
\renewcommand\arraystretch{1.15}
\begin{tabular}{|c|c|c|c|}
\hline
& NLO theory & NNLO & NNLO (no errors) \\
\hline 
 ATLAS, $R_{\rm low}$& 215.3 & 172.3 & 179.1 \\
 \hline
 ATLAS, $R_{\rm high}$ & 159.2 & 149.8 & 153.5 \\
 \hline 
 CMS, $R_{\rm low}$& 194.2 & 177.8  & 182.8\\
 \hline
 CMS, $R_{\rm high}$ & 198.5 & 182.3 & 188.8 \\
\hline
\end{tabular}
\caption{The $\chi^2$ for the combined NNLO fit to the ATLAS and CMS 7 TeV jet data, excluding and including the calculated NNLO K--factors, and excluding the errors associated with the polynomial fit to the K--factors. The $p_\perp^{\rm jet}$ factorization/renormalization scale is taken.}
\label{tab:combnnlonk}
\end{center}
\end{table}

\section{Impact of LHC jet data on PDFs}\label{sec:PDF}

\subsection{Central values}

\begin{figure}
\begin{center}
\includegraphics[scale=0.66]{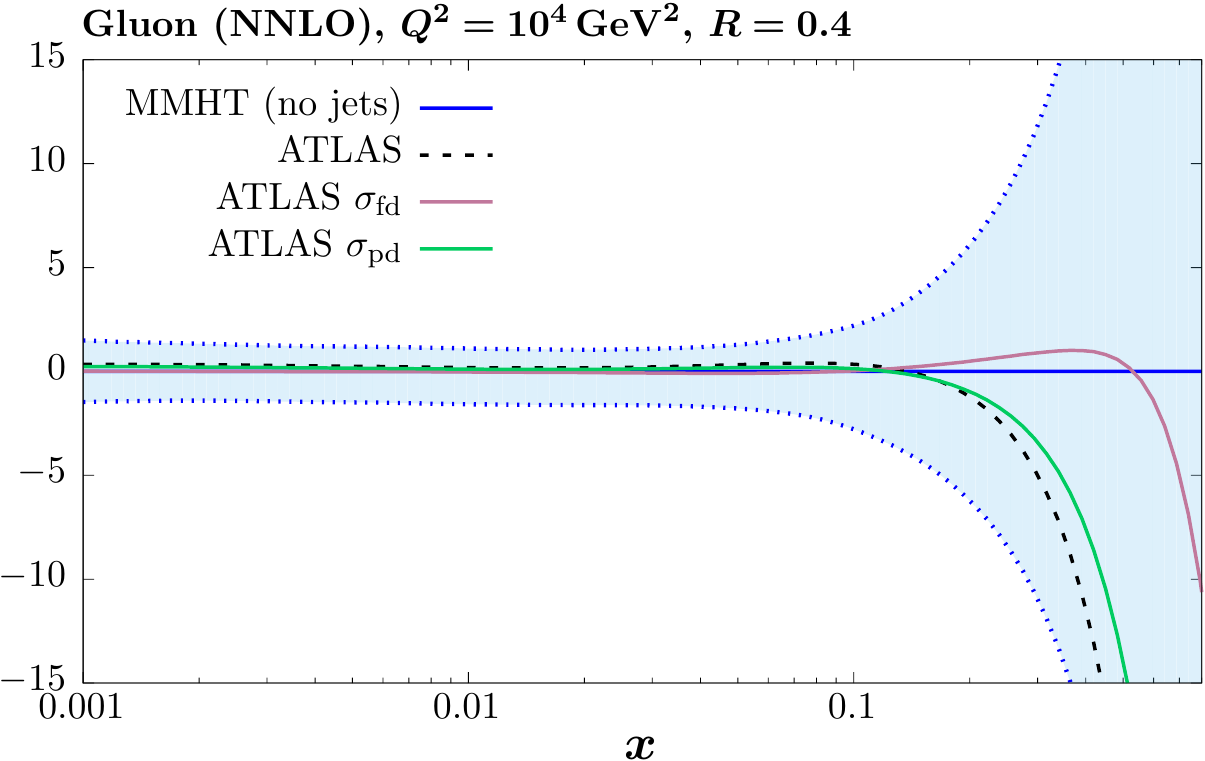}
\includegraphics[scale=0.66]{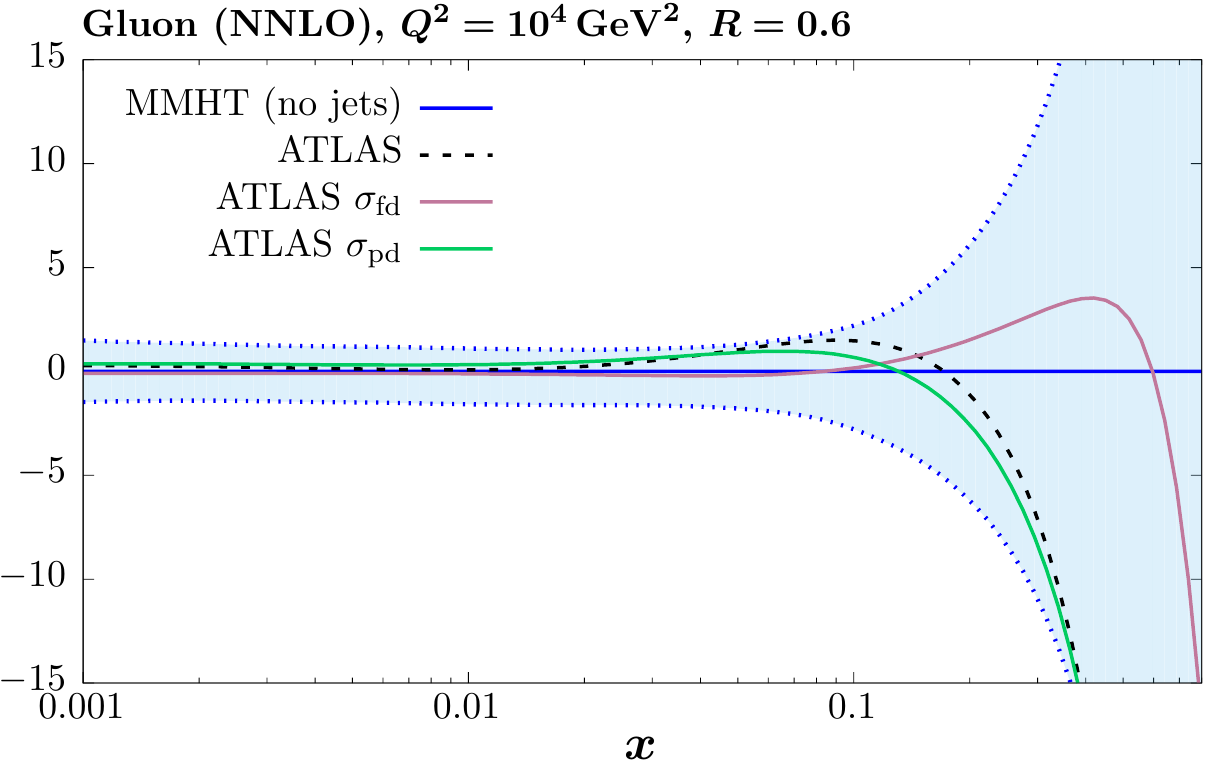}
\caption{The impact on the gluon PDF at NNLO of the ATLAS 7 TeV jet data~\cite{Aad:2014vwa}, including two alternative treatments of the correlated systematic errors described in the text. The percentage difference in comparison to the baseline fit, with no jet data included, is shown. The jet radii $R=0.4$ (0.6) ATLAS data is used in the left (right) plot.}
\label{fig:atlas}
\end{center}
\end{figure}

\begin{figure}
\begin{center}
\includegraphics[scale=0.66]{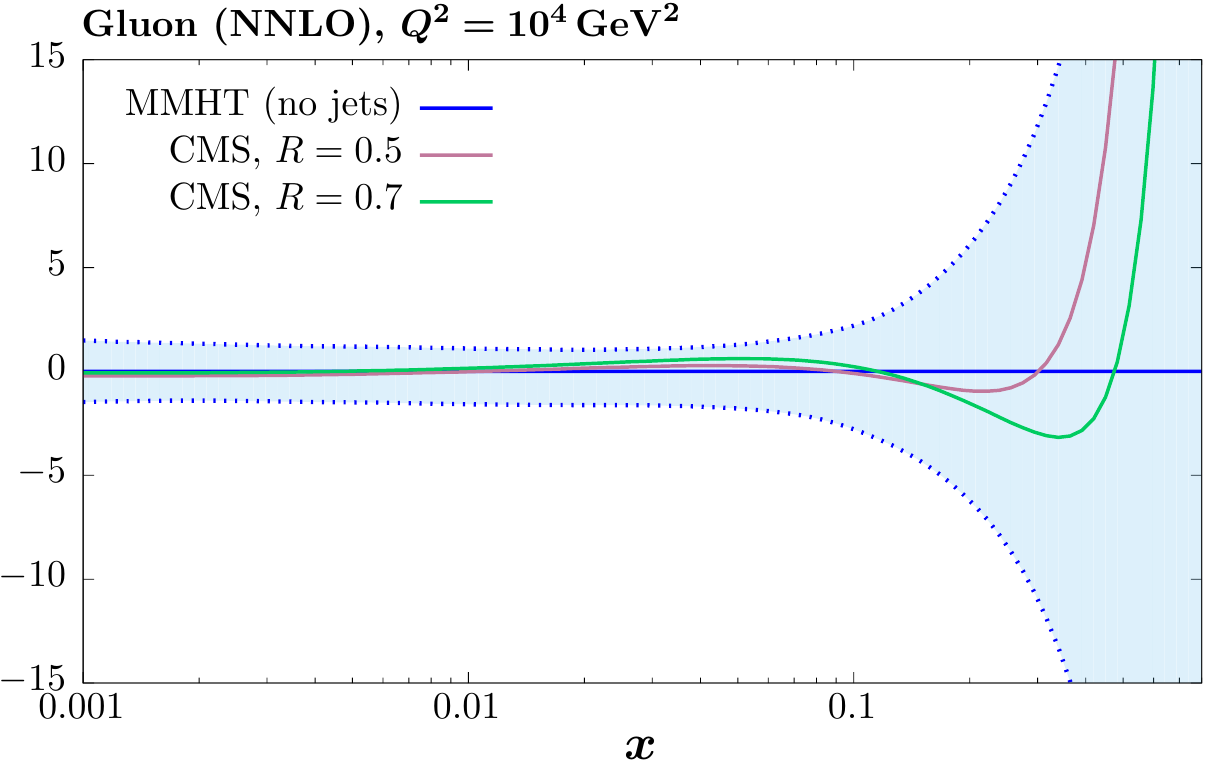}
\caption{The impact on the gluon PDF at NNLO of the CMS 7 TeV jet data~\cite{Chatrchyan:2014gia}, for two value of jet radii. The percentage difference in comparison to the baseline fit, with no jet data included, is shown.}
\label{fig:cms}
\end{center}
\end{figure} 

In this section we investigate the impact of including the jet data on the PDFs. We concentrate on the gluon PDF, as the effect on all quark PDF combinations is significantly smaller. 
In Fig.~\ref{fig:atlas} we show the impact of including the ATLAS jet data only in the fit, in comparison to the MMHT baseline described in the previous section (i.e. with Tevatron jet data omitted). We show the result with $R=0.4$ (0.6) in the left (right) figure, with the different treatments of the systematic errors described above. Only the comparison at NNLO is shown here, leaving the comparison to NLO for the combined fit to be presented below. Unless otherwise stated, in what follows we take $p_\perp^{\rm jet}$ as the choice of scale.

We can see that for both jet radii, despite leading to significantly different fit qualities, the partial decorrelation and default error treatments in fact result in quite similar fits for the gluon PDF, with some softening observed at high $x$. On the other hand, the full decorrelation of systematic uncertainties leads to a gluon that is qualitatively different, being much less soft at high $x$, although still consistent within PDF uncertainties. This is perhaps not surprising, as the systematic shifts we determine by profiling with respect to the various correlated uncertainties in (\ref{eq:chicor}) have a physical interpretation, giving us the best fit values of the various experimental parameters and a corresponding best fit measurement that is shifted with respect to the default. By treating these sources of uncertainty as uncorrelated across rapidity bins, this connection is largely lost, and in effect an imperfect measurement that is systematically different may be fit. The central value of the extracted gluon may then vary quite significantly. This effect is indeed observed in Fig.~\ref{fig:atlas}. Given these results, in what follows we will simply apply our model of partial error decorrelation, although we note that in all cases the results are very similar when taking the default treatment.

It is interesting to observe in Fig.~\ref{fig:atlas} that the difference due to the choice of jet radius is relatively small, and much less than that due to the error treatment, although the higher $R=0.6$ choice leads to a somewhat softer gluon at high $x$. In Fig.~\ref{fig:cms} we show the result of the NNLO fit, including the CMS jet data only, for both jet radii. Here, the impact on the gluon is relatively flat out to quite high $x$, where some hardening is observed, albeit within the large PDF uncertainties in this region. As with the ATLAS data, the larger choice of jet radius leads to some softening in the gluon in comparison to the lower choice.

\begin{figure}
\begin{center}
\includegraphics[scale=0.66]{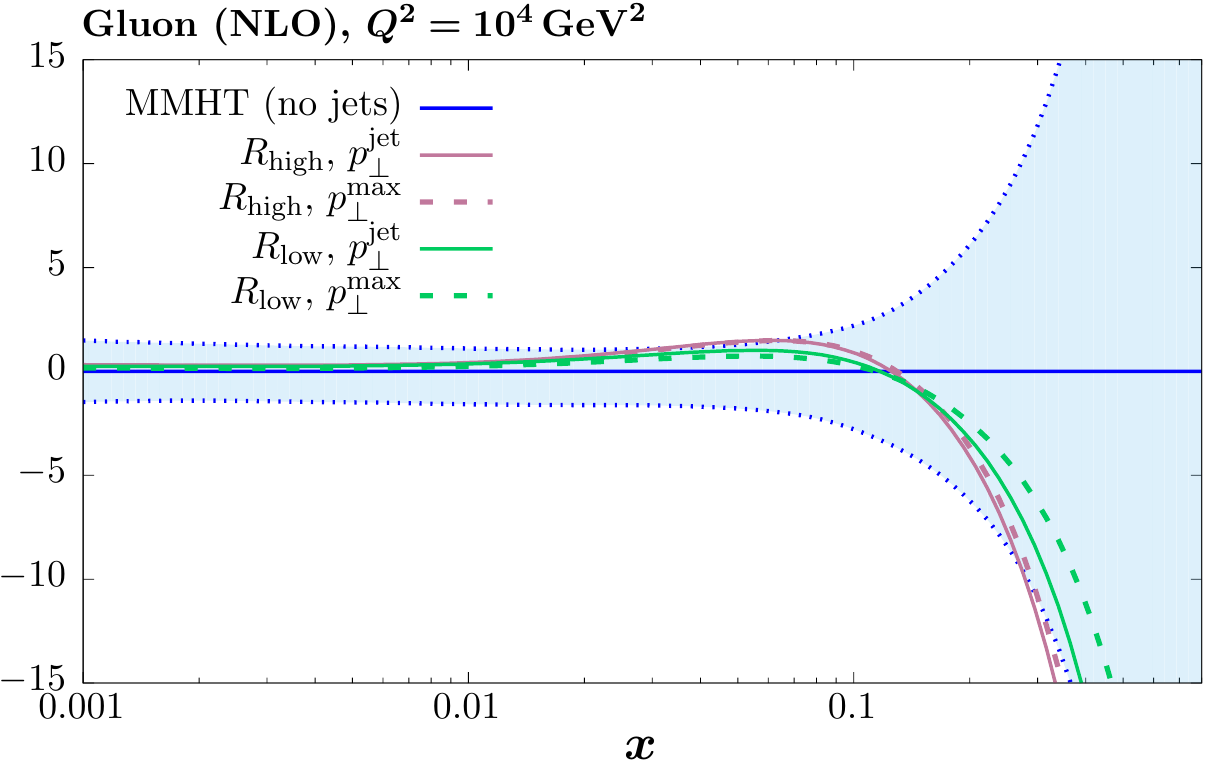}
\includegraphics[scale=0.66]{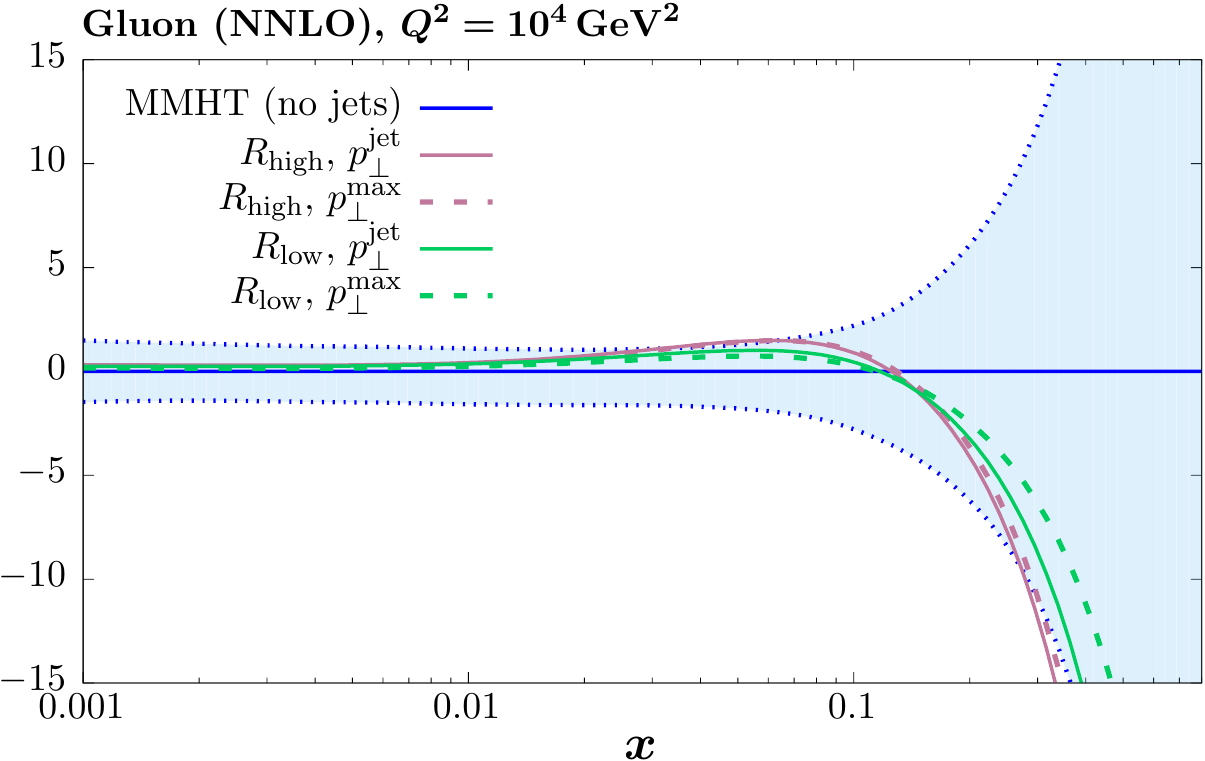}
\caption{The impact of the ATLAS~\cite{Aad:2014vwa} and CMS~\cite{Chatrchyan:2014gia} 7 TeV jet data on the gluon PDF at NLO (left) and NNLO (right). The percentage difference in comparison to the baseline fit, with no jet data included, is shown. Results are given for `low' and `high' jet radii described in the text, and for two choices of the factorization scale.}
\label{fig:comb}
\end{center}
\end{figure} 

\begin{figure}
\begin{center}
\includegraphics[scale=0.4]{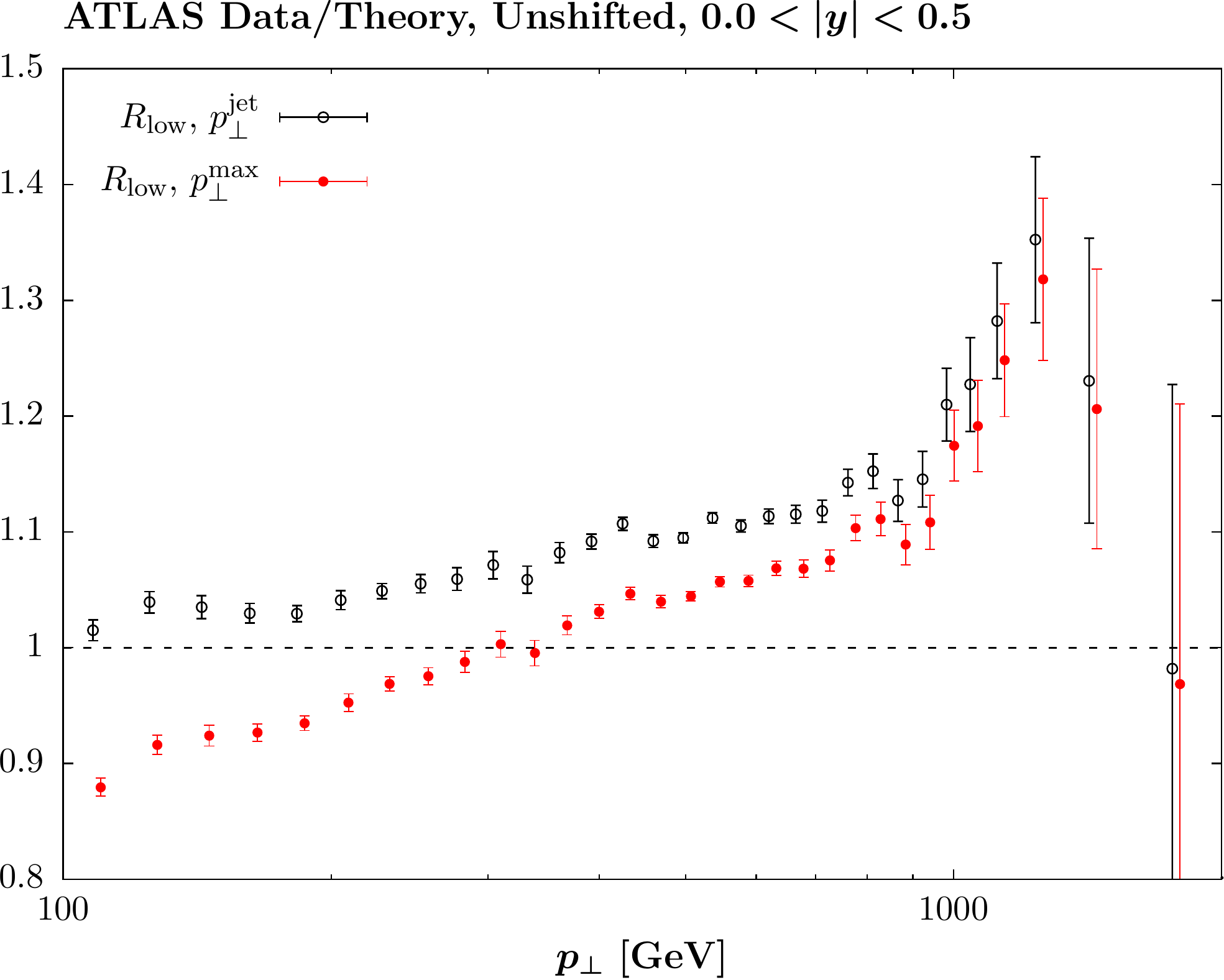}\quad
\includegraphics[scale=0.4]{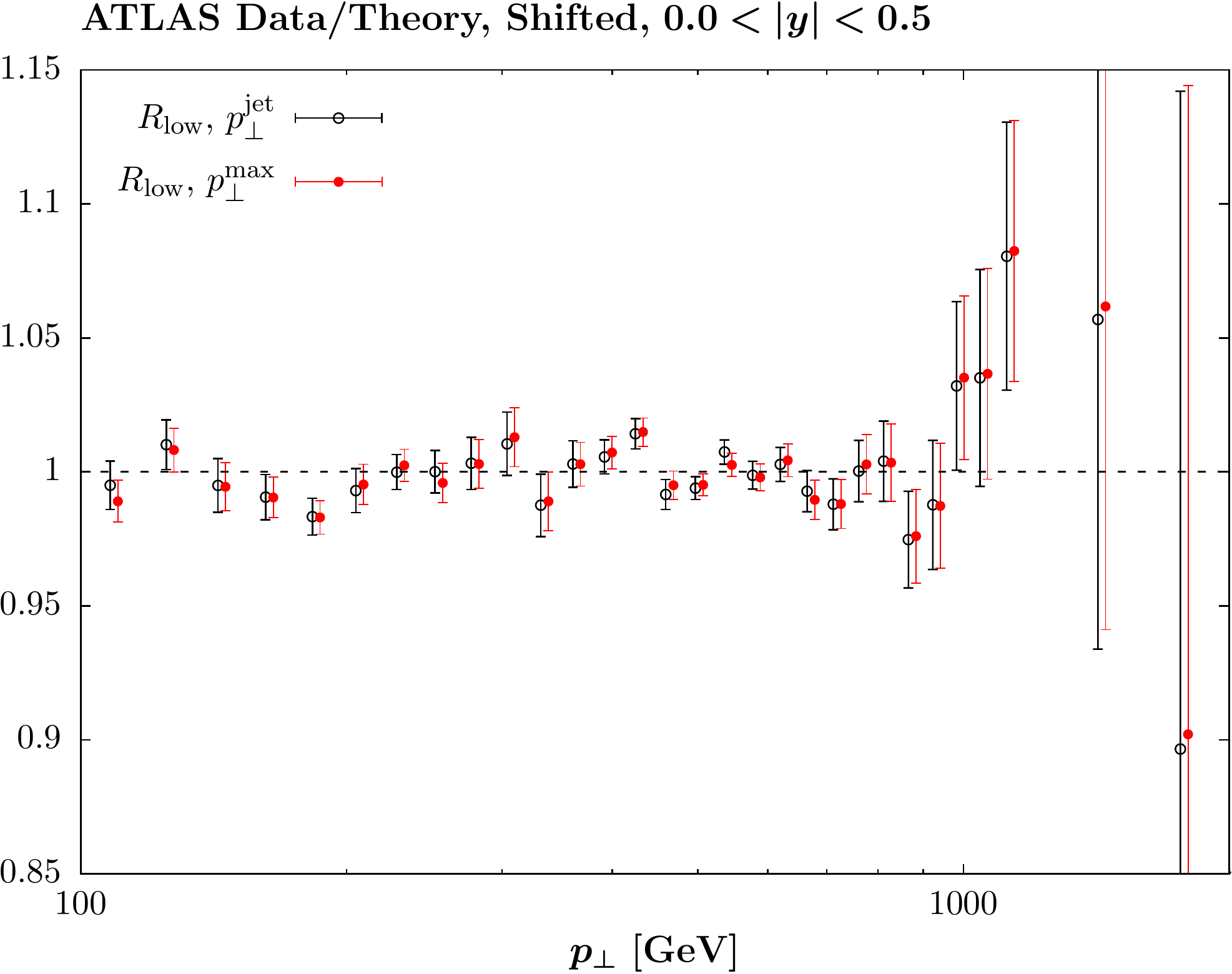}\vspace{1cm}
\includegraphics[scale=0.4]{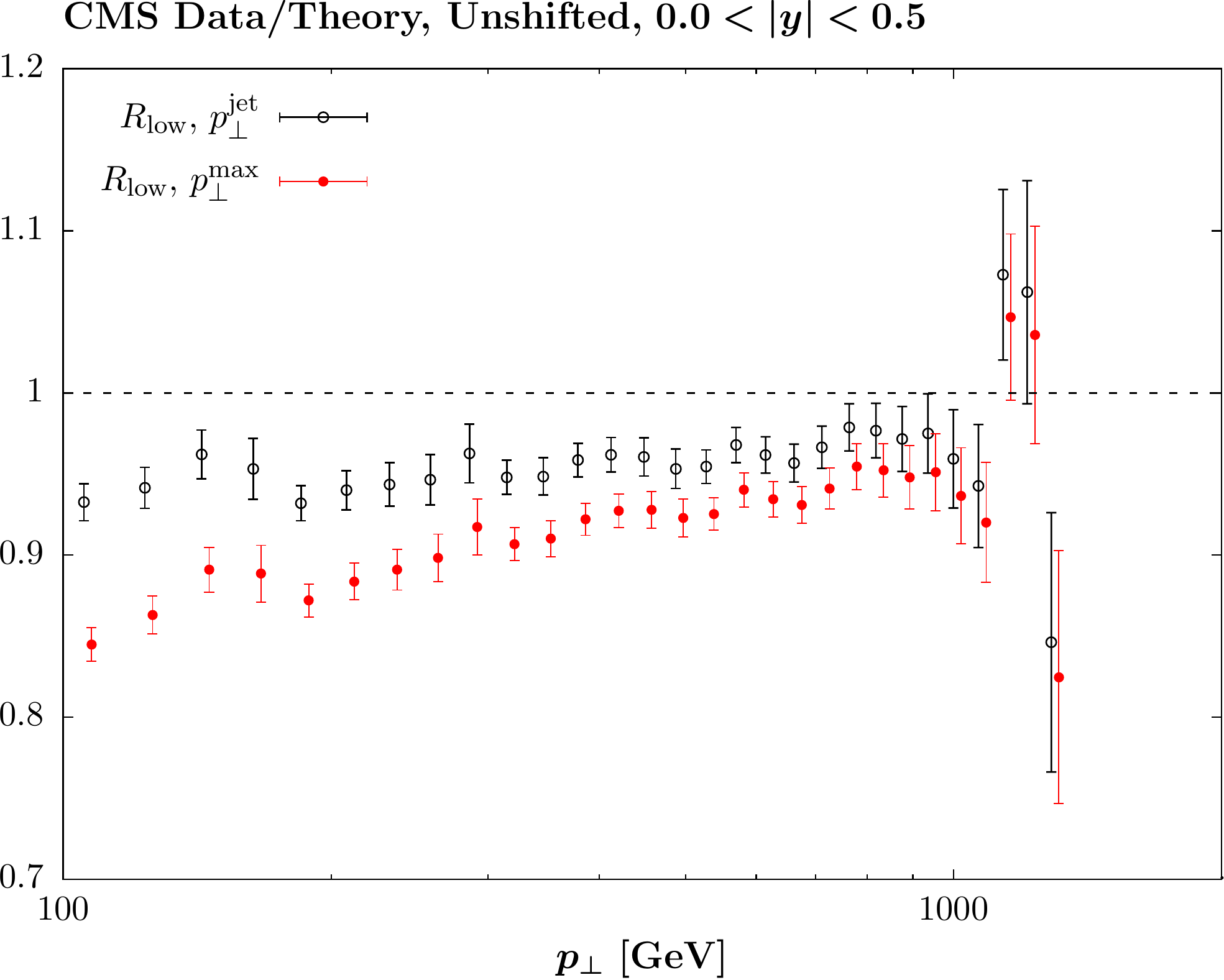}\quad
\includegraphics[scale=0.4]{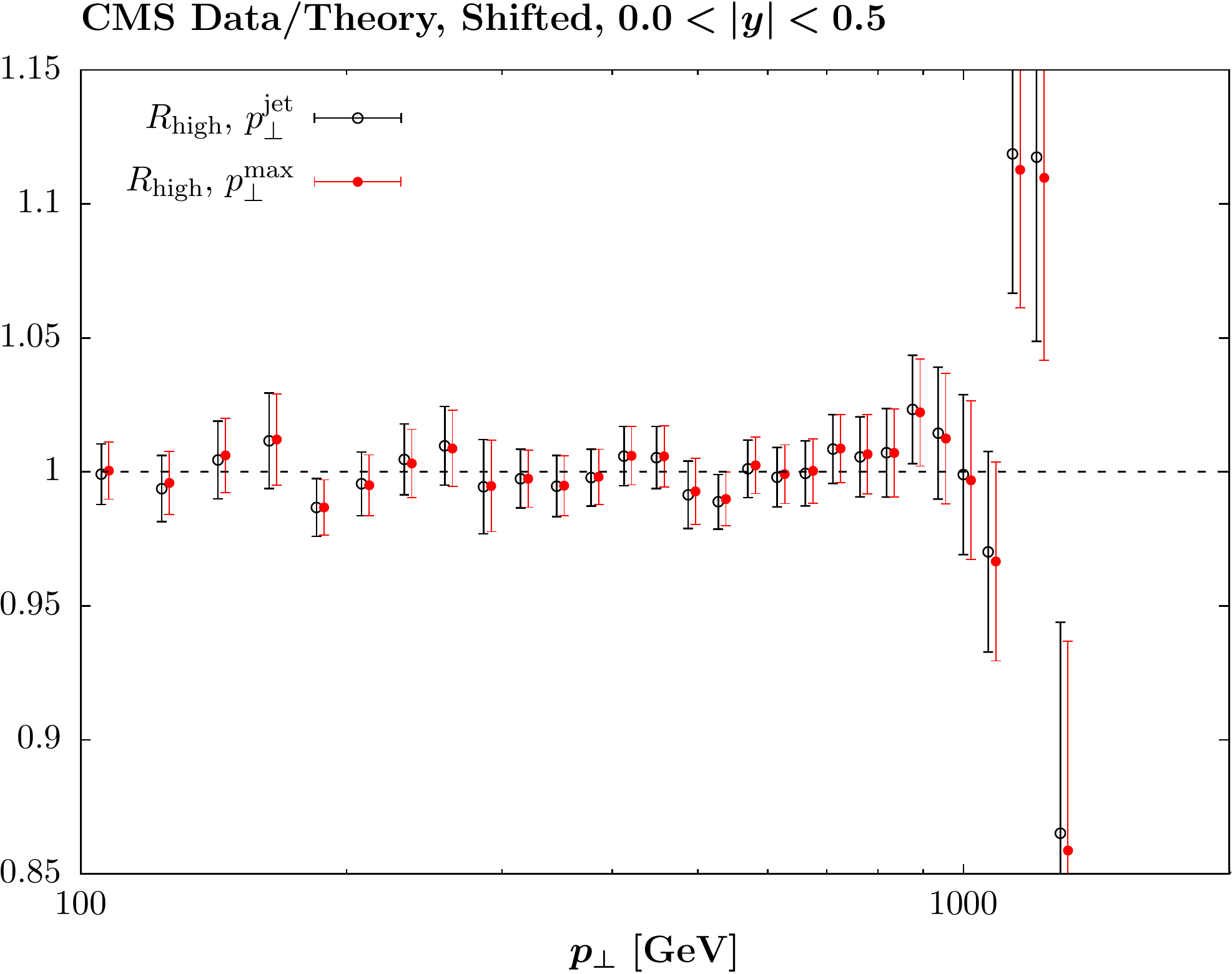}
\caption{Data/theory for the NNLO fit including the combined ATLAS and CMS data with the lower $R$ radius choice and for both choices of scale. The comparison in the central rapidity bin for the ATLAS (CMS) data is shown in the upper (lower) plots. The result before and after the inclusion of the shifts due to the correlated systematic errors is shown in the left and right hand plots, respectively. Statistical errors only are shown.}
\label{fig:scdata}
\end{center}
\end{figure} 

\begin{figure}
\begin{center}
\includegraphics[scale=0.4]{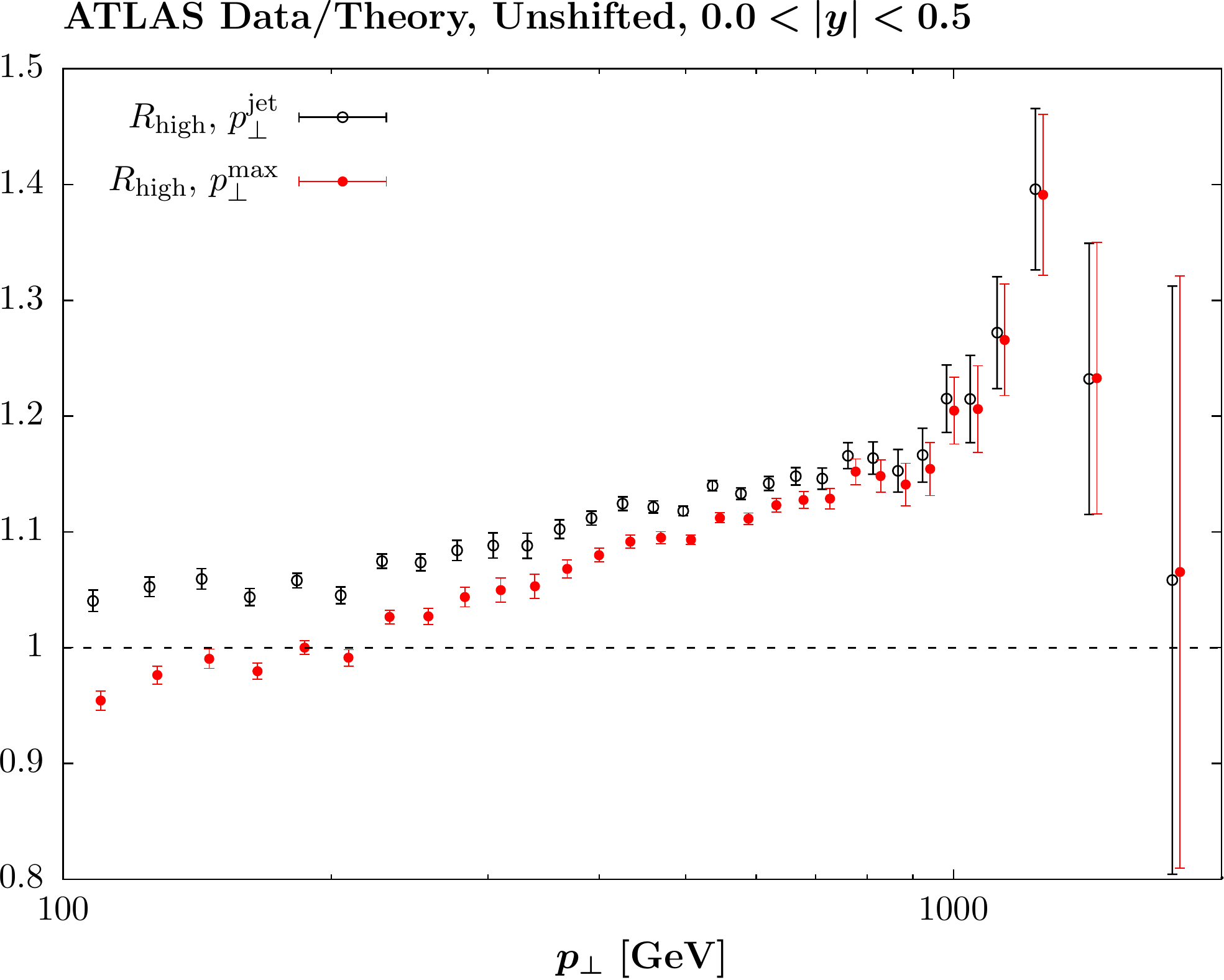}\quad
\includegraphics[scale=0.4]{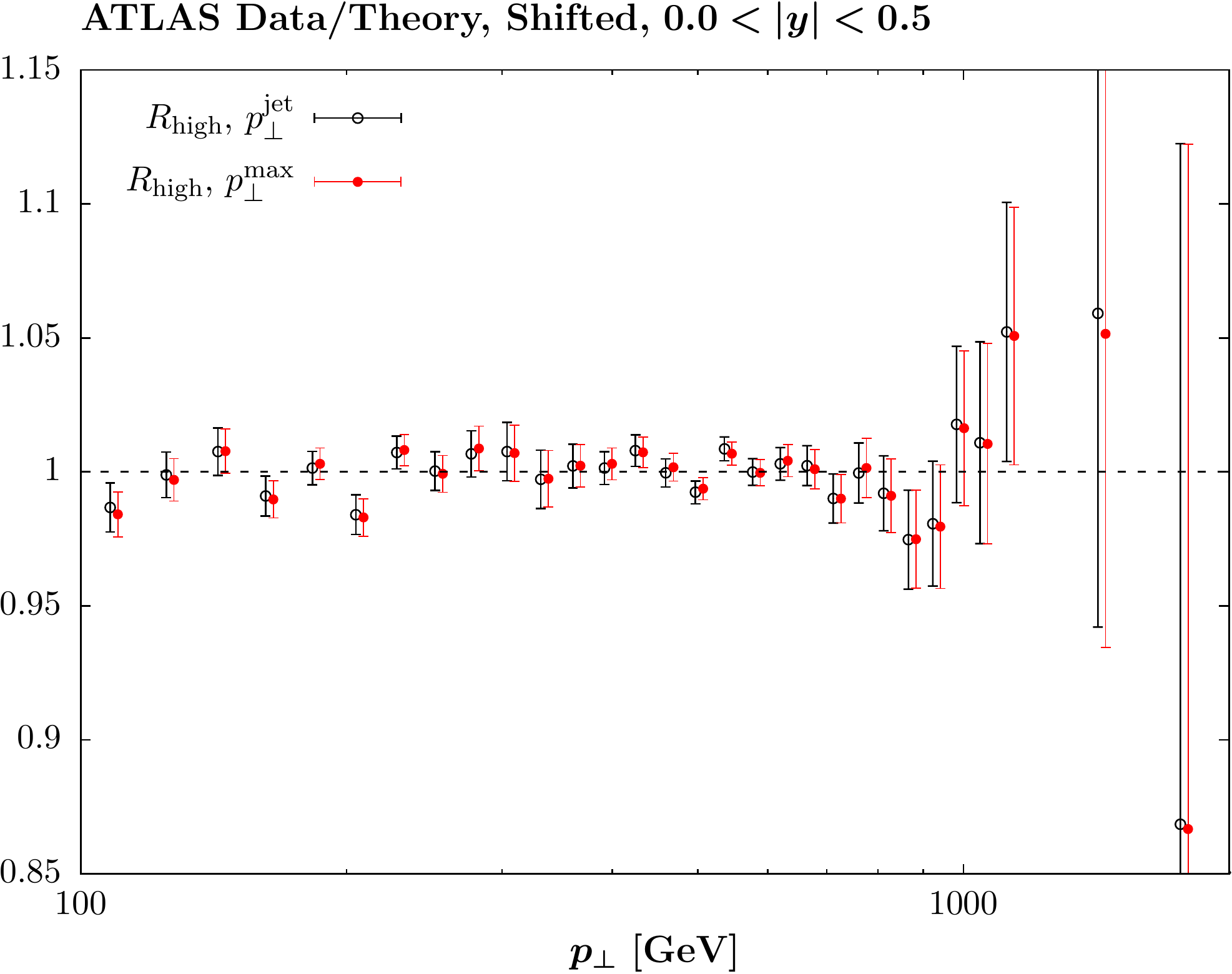}
\includegraphics[scale=0.4]{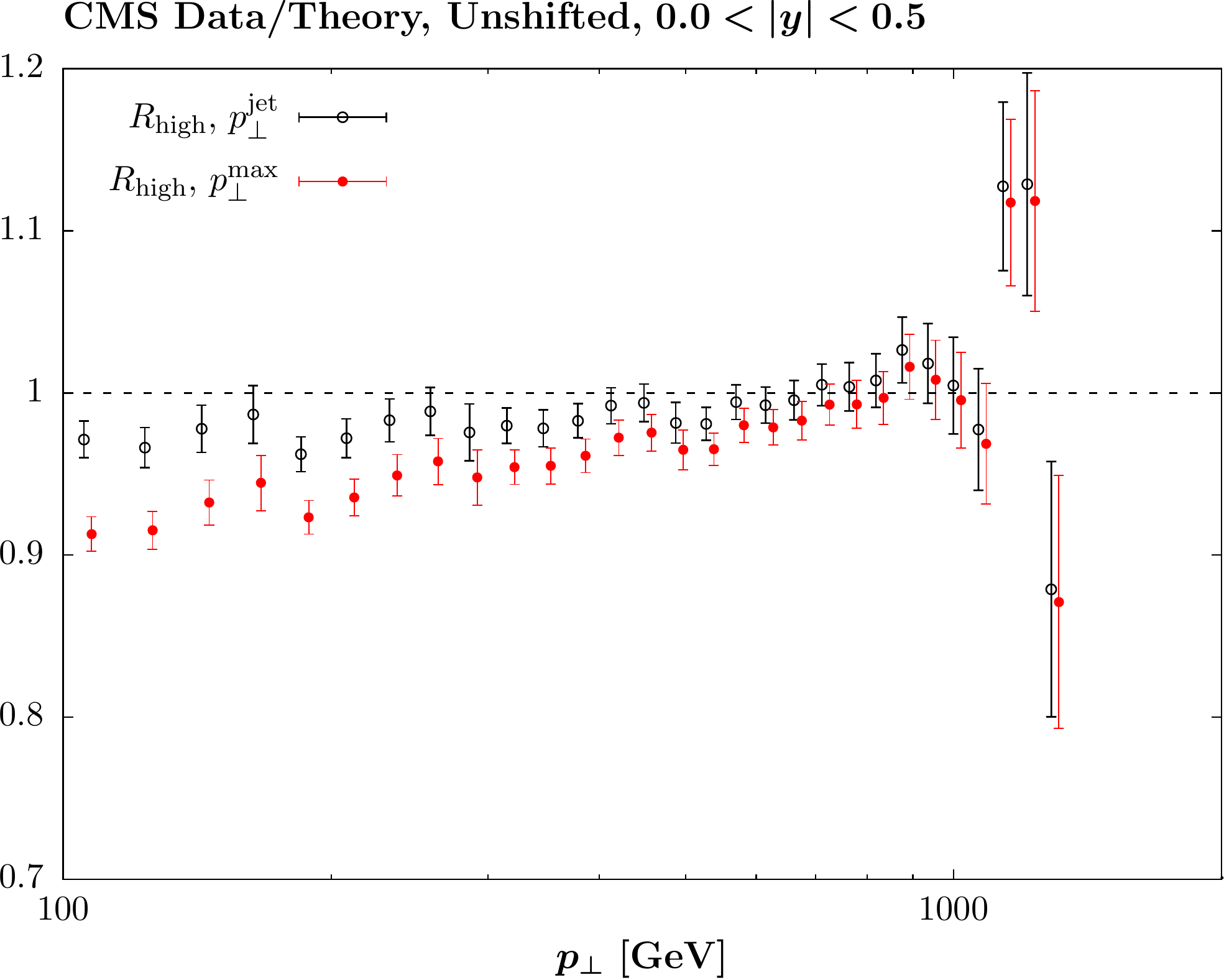}\quad
\includegraphics[scale=0.4]{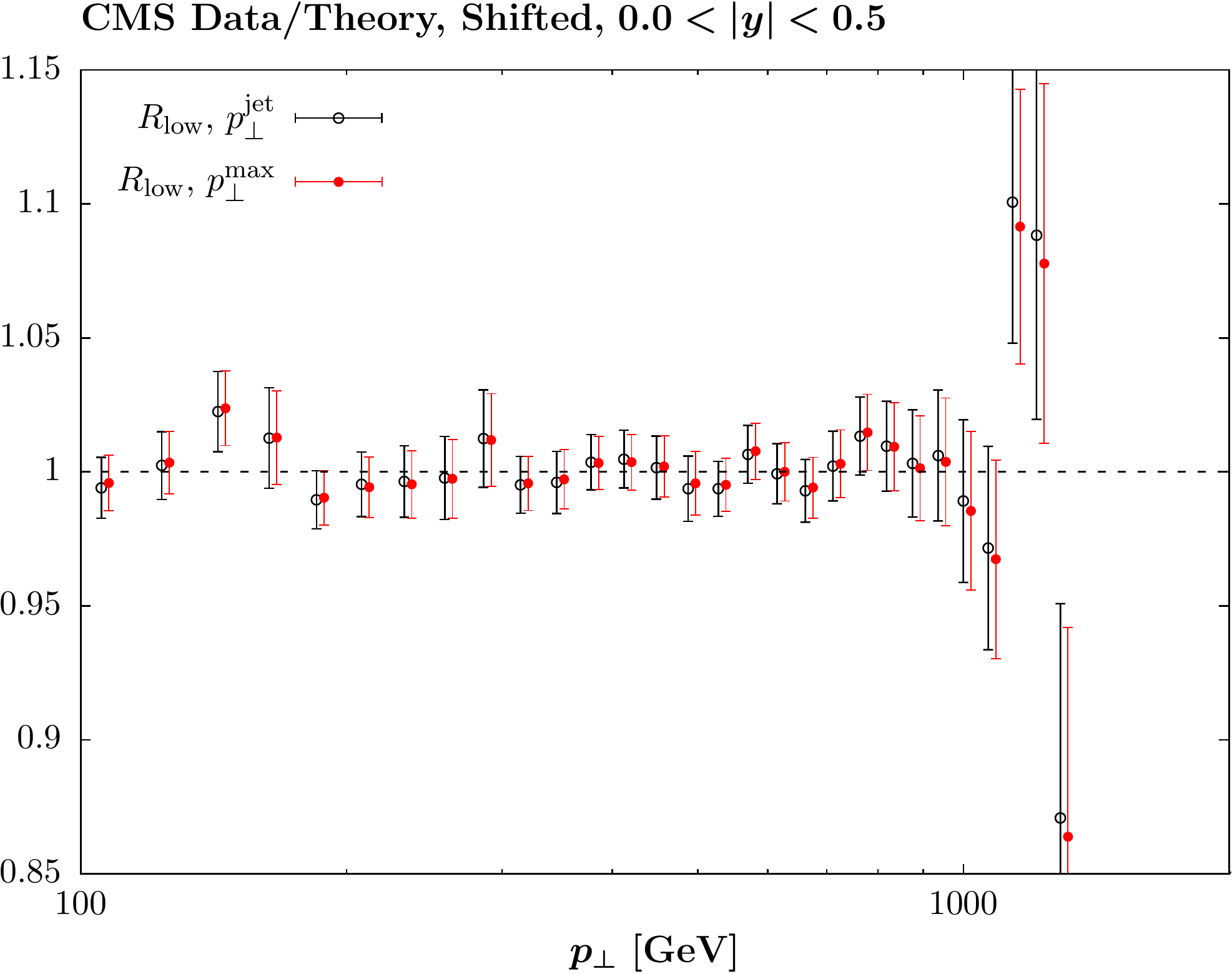}
\caption{As in Fig.~\ref{fig:scdata}, but with the higher $R$ radius choice.}
\label{fig:scdatarh}
\end{center}
\end{figure} 

In Fig.~\ref{fig:comb} we now consider the effect of combined fit to the ATLAS and CMS jet on the gluon. As mentioned above, we take the partial decorrelated treatment of the ATLAS jet data in what follows. We show results for low and high jet radii, i.e. with $R=0.4$ (0.5) and $R=0.6$ (0.7) for the ATLAS (CMS) data, respectively. We also show the effect of taking the $p_\perp^{\rm max}$ scale choice in comparison to $p_\perp^{\rm jet}$. The result at NLO (NNLO) is shown in the left (right) panel. The impact of the scale choice on the gluon is quite small, of the same order of or less than that due to the choice of jet radius, although here the difference for the combined fit is also not dramatic. This is not necessarily to be expected, as the difference between the scale choices in the underlying theory prediction is not negligible.  In addition, while the qualitative trend in the NLO and NNLO fits is similar, the latter leads to a somewhat softer gluon, which even lies somewhat outside the baseline PDF uncertainty band for the higher jet radius. We can also see that the gluon that results from the combined fit lies closer to the result from the ATLAS then the CMS fit, although these are all consistent within PDF uncertainties. This is consistent with the somewhat larger deterioration observed in the fit quality for the CMS--only case in comparison to the combined fit.

\begin{figure}
\begin{center}
\includegraphics[scale=0.66]{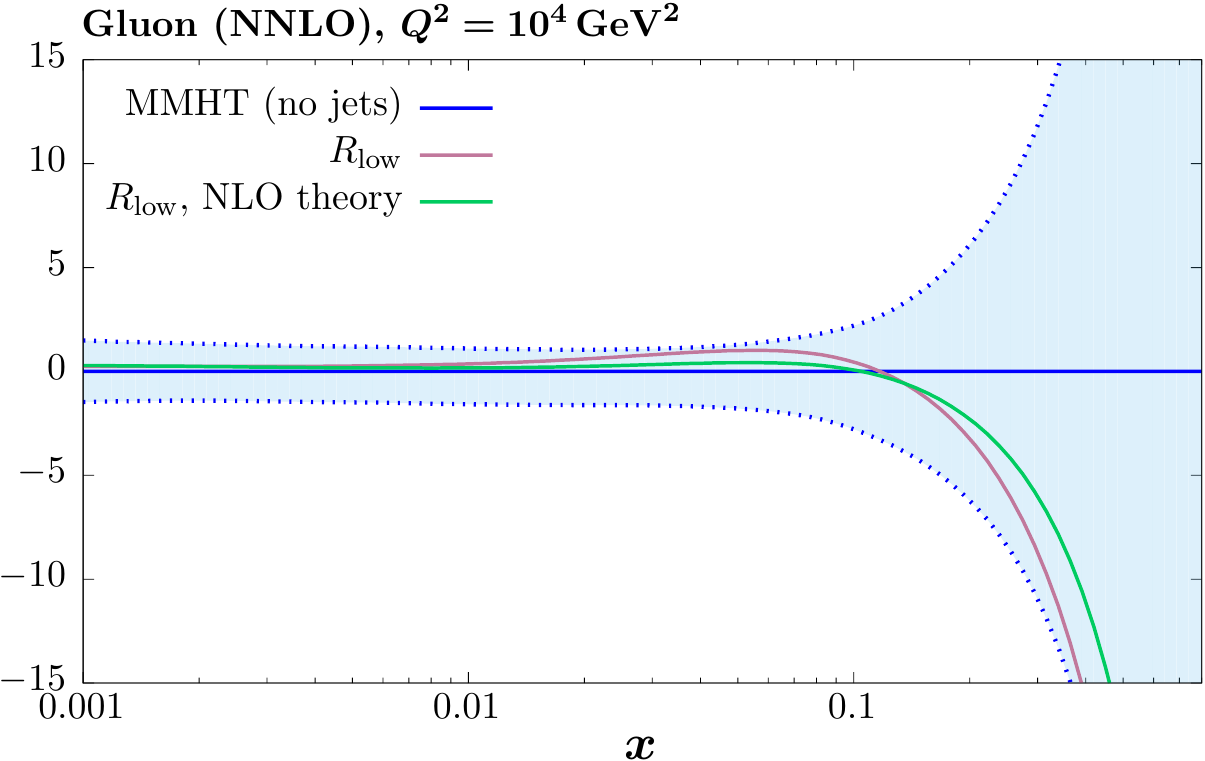}
\includegraphics[scale=0.66]{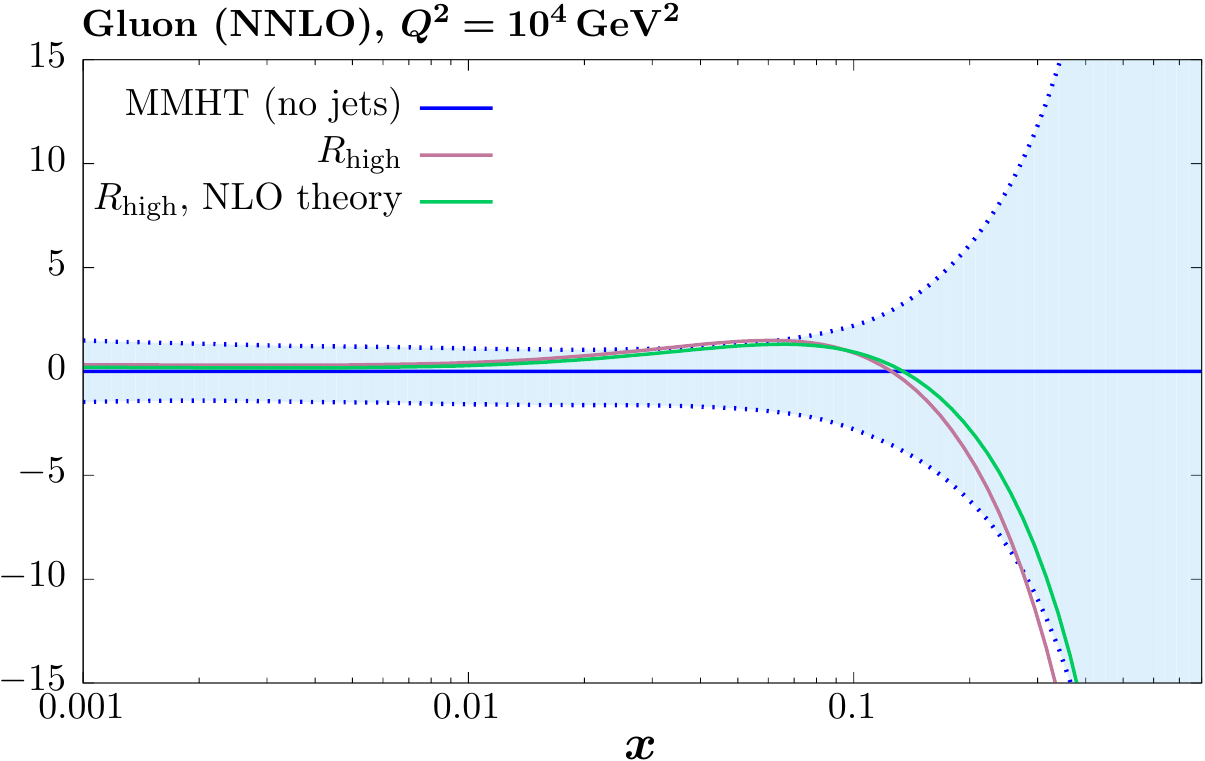}
\caption{The impact on the gluon PDF at NNLO of the ATLAS~\cite{Aad:2014vwa} and CMS~\cite{Chatrchyan:2014gia} 7 TeV jet data. The percentage difference in comparison to the baseline fit, with no jet data included, is shown. Results are shown for the default NNLO fit, and to the same fit but using the NLO theory, i.e. with the NNLO K-factors omitted. The left (right) plots correspond to the `low' and `high' jet radii described in the text.}
\label{fig:combnk}
\end{center}
\end{figure} 

Thus, to summarise the effect of the LHC data and the accompanying theory improvements, in all cases we observe some relative stability in the overall trend of the extracted gluon, with the smallest differences being due to the choice of scale, followed by the choice of jet radius and finally the NLO vs. NNLO difference being largest. We study this result in more detail below.

To investigate the effect of scale choice further, in Figs.~\ref{fig:scdata} and~\ref{fig:scdatarh} we show the data/theory for both choices of scale, with lower and higher $R$ choice, respectively. We show results for the both the ATLAS and CMS data, in the central rapidity bin (although similar results are seen at other rapidities). In the left hand plots we show the results prior to including the systematic shift in the correlated errors. We concentrate our discussion below on the lower $R$ choice shown in Fig.~\ref{fig:scdata}, as here  difference with respect to the two scale choices is more pronounced, but similar conclusions hold for the higher choice, shown in Fig.~\ref{fig:scdatarh}. An approximately $10\%$ difference is observable at lower $p_\perp$, with the $p_\perp^{\rm max}$ choice leading to the larger result, consistent with the findings in~\cite{Currie:2017ctp}. We can see that in both cases the description of the data is poor, highlighting the importance of the systematic experimental uncertainties. However, once the data is allowed to shift by these errors this difference largely disappears, and good description of the data is achieved in all cases. The shift is somewhat larger for the $p_\perp^{\rm max}$ case, with a $\sim 5$ (11) point increase in the $\chi^2$ due the shift penalty found for the ATLAS (CMS) data, while for the higher $R$ choice the overall shift penalty is only marginally increased, by 2 points. These findings are consistent with the trends found in Table~\ref{tab:combnnlo}. From Fig.~\ref{fig:comb} we can see that these results translate into a relative, although not complete, stability in the predicted PDFs. With a further reduction in the size of the systematic experimental uncertainties the difference may on the other hand become more pronounced, but no significant effect is observed with the 7 TeV data sets.

To investigate the impact of the NNLO corrections further, in Fig.~\ref{fig:combnk} we again show the NNLO gluon resulting from the combined fit, but including the result with the NLO theory applied, i.e. excluding the NNLO K--factors in the fit. In the left (right) panel we show the result with the low (high) choice of jet radius. This therefore shows the impact of the new NNLO theory calculation on the gluon. We can see that in both cases the effect is reasonably small, but not negligible, leading to some additional softening in the gluon at high $x$. Indeed, for the high jet radius choice, the inclusion of the NNLO theory leads to a central value at high $x$ which lies somewhat outside the uncertainty band of the baseline fit. The effect  of using $p_\perp^{\rm max}$ instead as the scale choice is similar. 

\begin{figure}
\begin{center}
\includegraphics[scale=0.66]{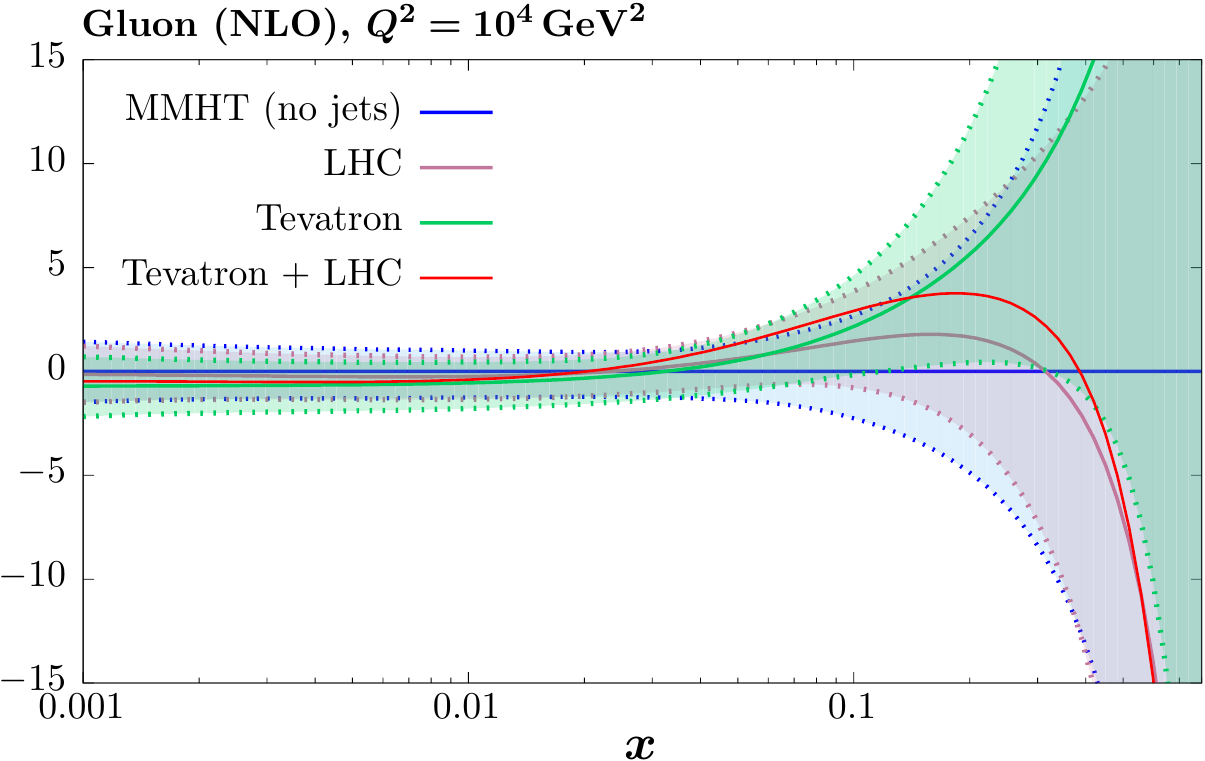}
\includegraphics[scale=0.66]{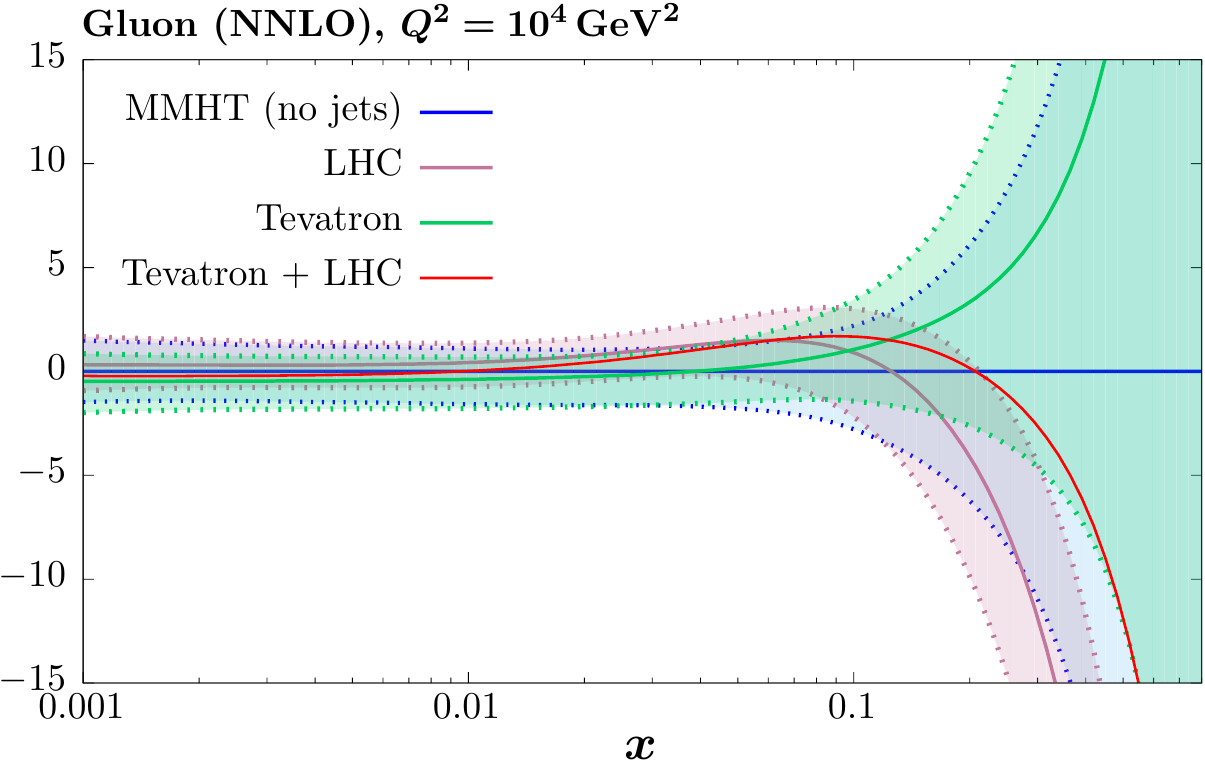}
\caption{The impact on the gluon PDF when fitting the ATLAS~\cite{Aad:2014vwa} and CMS~\cite{Chatrchyan:2014gia} 7 TeV jet data and Tevatron data~\cite{Abazov:2011vi,Abulencia:2007ez} individually, as well as including all datasets within the fit. For the LHC case $R_{\rm high}$ and $p_\perp^{\rm jet}$ are taken. The NLO (NNLO) results are shown in the left (right) plots.}
\label{fig:combtevlhc}
\end{center}
\end{figure}

Finally, we consider the impact of including the Tevatron jet data~\cite{Abazov:2011vi,Abulencia:2007ez} on the NLO and NNLO fits. As discussed above, for the NNLO case the full calculation is not yet publicly available, and so continue to apply the threshold corrections of~\cite{Kidonakis:2000gi}. We show in Fig.~\ref{fig:combtevlhc} the result of including the Tevatron data alone, as well as the Tevatron and LHC data. The fit to the LHC jet data (in all cases with $R_{\rm high}$ and $p_\perp^{\rm jet}$) is also shown for comparison. As the NLO and NNLO cases are qualitatively quite similar, we will only discuss the NNLO fit below. For the fit to the Tevatron data, an increase in the central gluon at higher $x$ is observed, consistent with its impact in the MSTW08 fit~\cite{Martin:2009iq}. This is in contrast to the LHC data, which we have seen prefers a softer gluon at higher $x$, although up to $x \gtrsim 0.3$ these are consistent within PDF errors. Nonetheless some tension is observed, and indeed when including both the Tevatron and LHC data into the fit, the description deteriorates by about 10 and 8 points in comparison to the individual fits for the LHC and Tevatron, respectively. The resultant gluon is somewhat harder at high $x$ than the LHC only fit, but still softer than the baseline. For clarity we do not include the PDF uncertainties in this case; these will be shown below. It will be interesting to see how this situation changes when the full NNLO corrections are included for the Tevatron predictions.

\subsection{PDF Uncertainties}

\begin{figure}
\begin{center}
\includegraphics[scale=0.66]{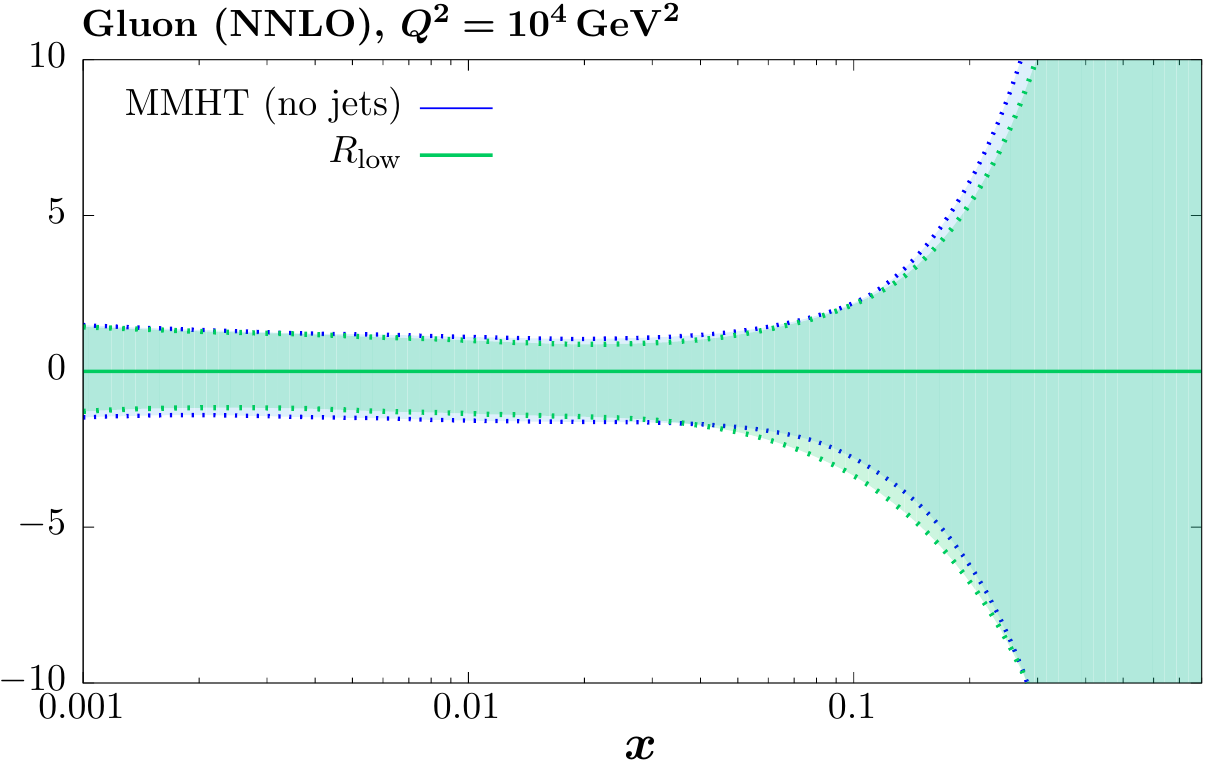}
\includegraphics[scale=0.66]{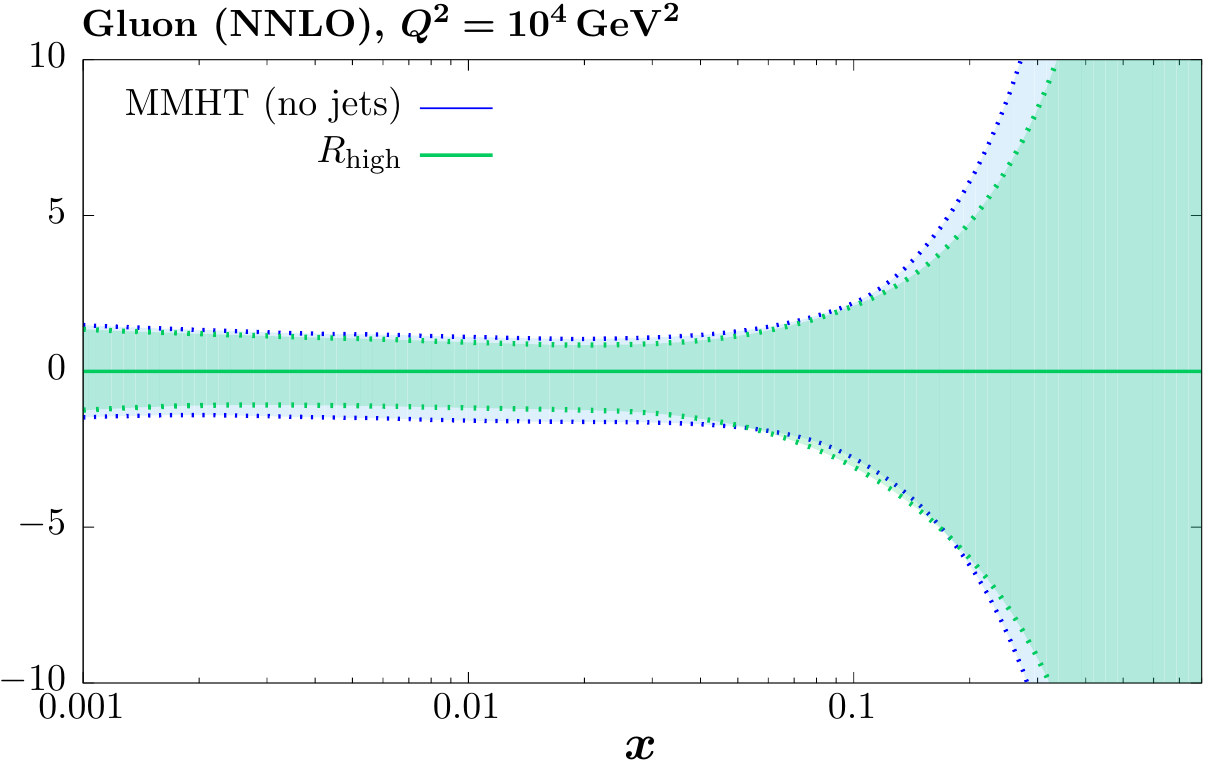}
\caption{The impact on the gluon PDF errors at NNLO of including the ATLAS~\cite{Aad:2014vwa} and CMS~\cite{Chatrchyan:2014gia} 7 TeV jet data in the global fit. The percentage errors at 68\% C.L. are shown, with the result of the baseline fit, with no jet data included, given for comparison.  The left (right) plots correspond to the for `low' and `high' jet radii described in the text.}
\label{fig:comberr}
\end{center}
\end{figure}

\begin{figure}
\begin{center}
\includegraphics[scale=0.66]{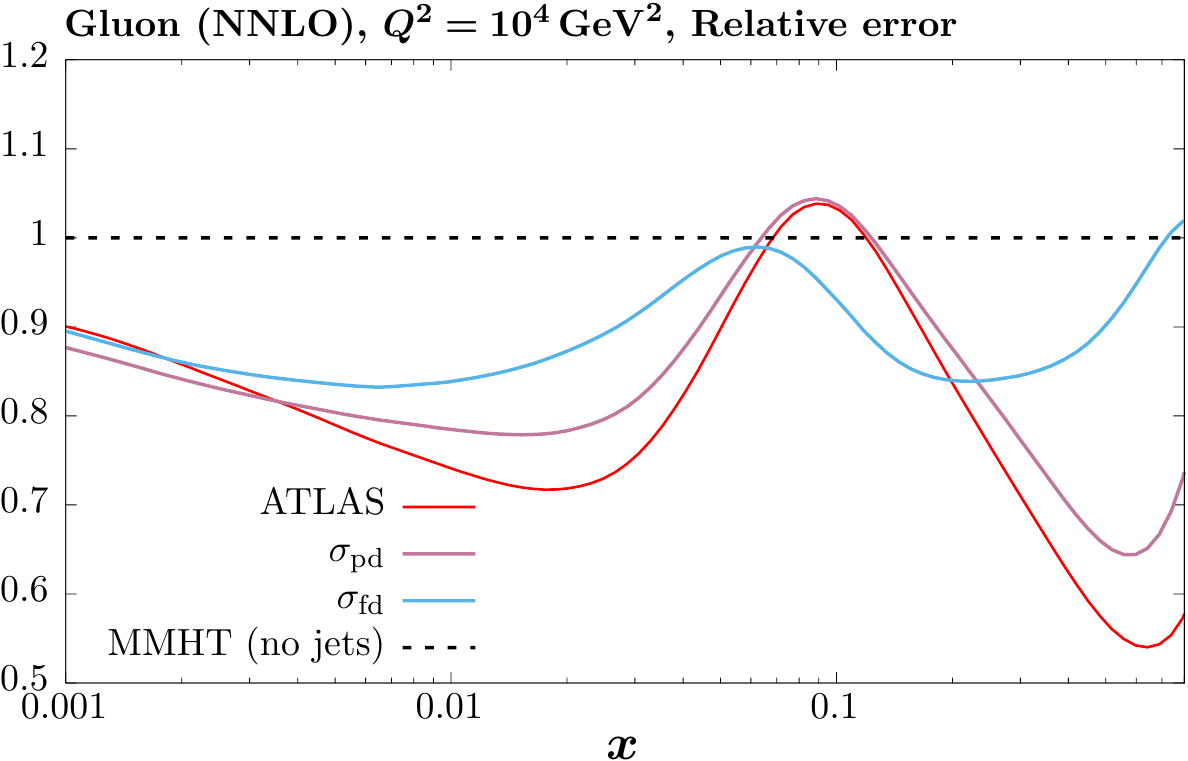}
\includegraphics[scale=0.66]{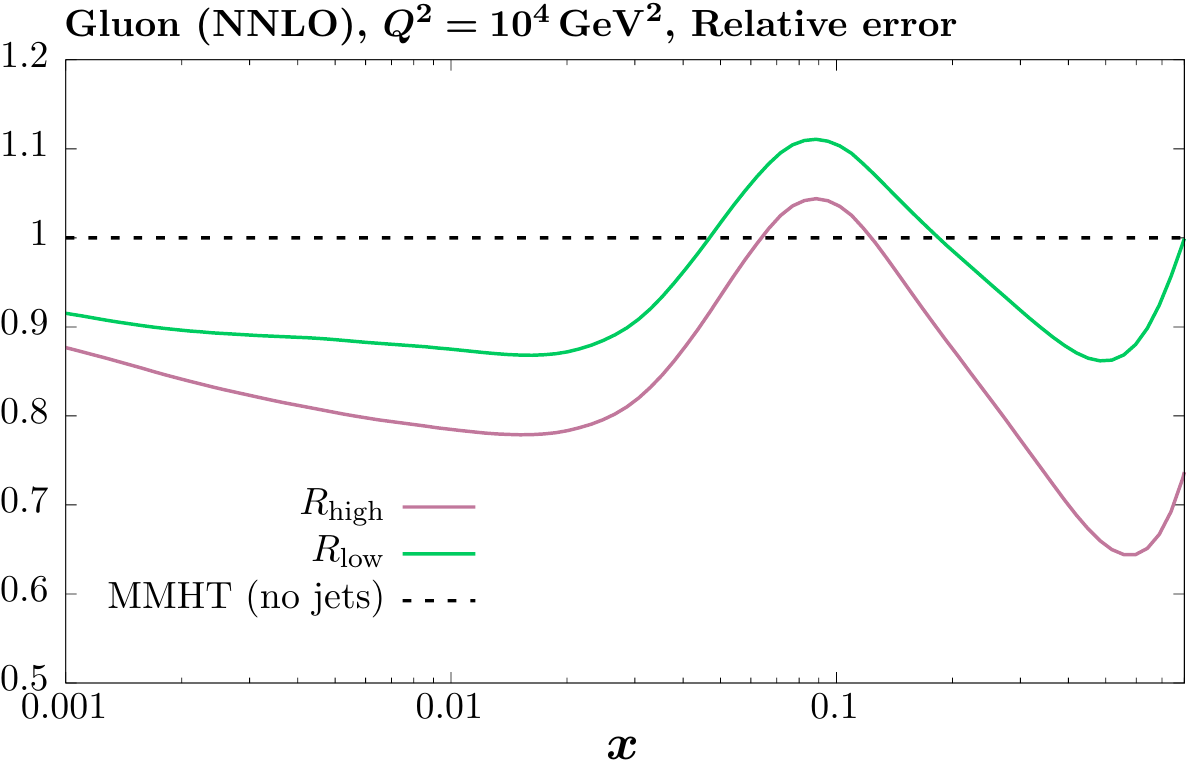}
\caption{The impact on the gluon PDF errors at NNLO of including the ATLAS~\cite{Aad:2014vwa} and CMS~\cite{Chatrchyan:2014gia} 7 TeV jet data in the global fit. The ratio of the 68\% C.L. errors to the baseline fit is shown for different choices of jet radius and treatment of systematic errors in the ATLAS case.}
\label{fig:combrelerr}
\end{center}
\end{figure} 

\begin{figure}
\begin{center}
\includegraphics[scale=0.66]{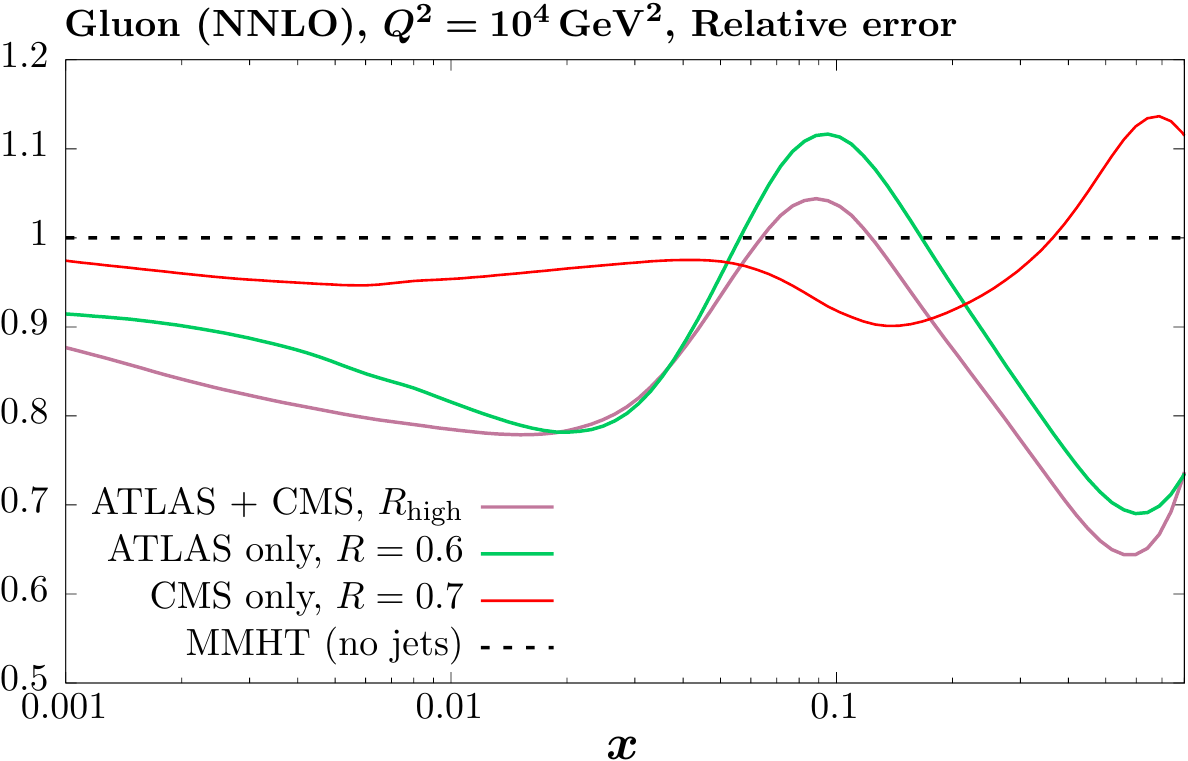}
\caption{The impact on the gluon PDF errors at NNLO of including the ATLAS~\cite{Aad:2014vwa} and CMS~\cite{Chatrchyan:2014gia} 7 TeV jet data in the global fit. The ratio of the 68\% C.L. errors to the baseline fit is shown. For the LHC case, $R_{\rm high}$ and $p_\perp^{\rm jet}$ are taken.}
\label{fig:combrelerr1}
\end{center}
\end{figure} 

\begin{figure}
\begin{center}
\includegraphics[scale=0.66]{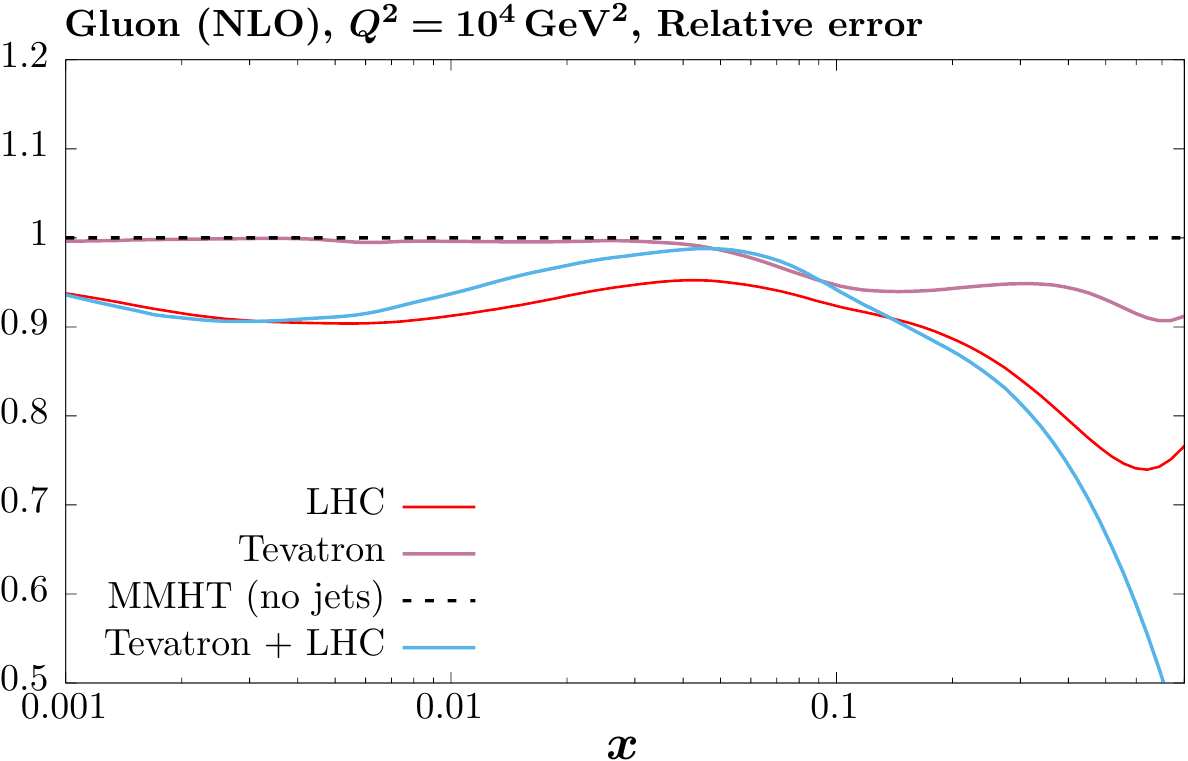}
\includegraphics[scale=0.66]{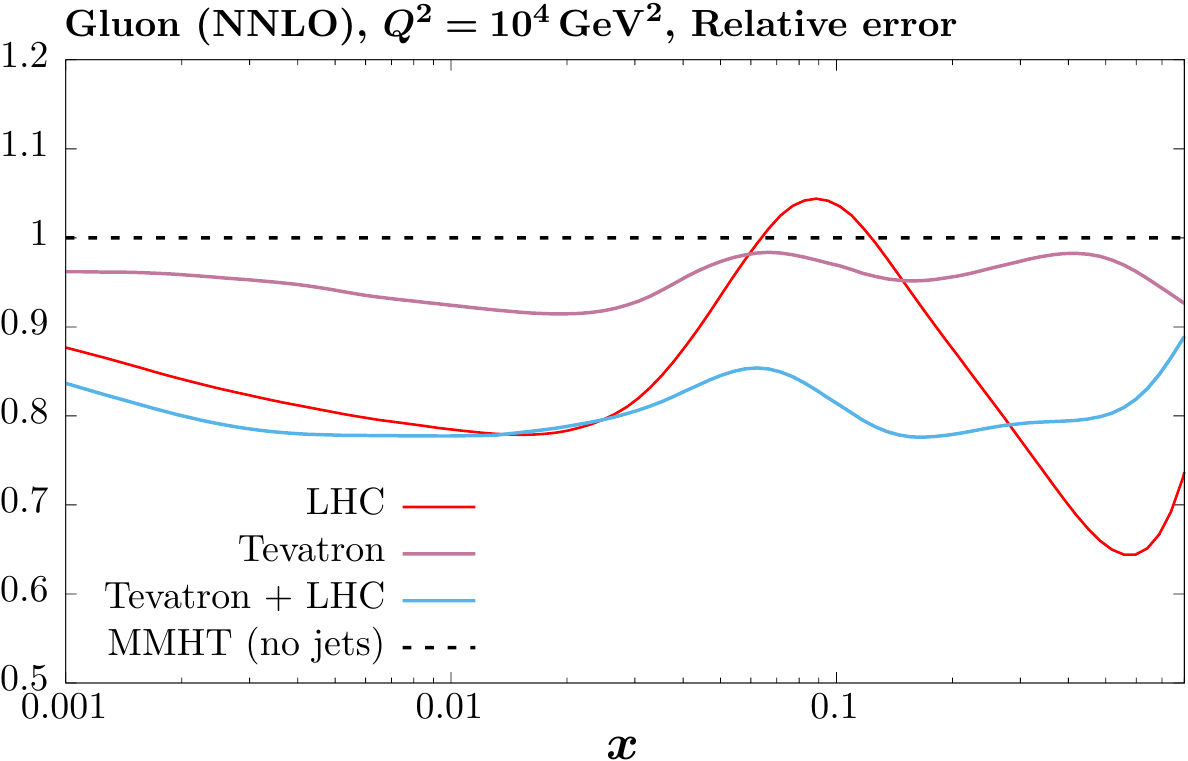}
\caption{The impact on the gluon PDF errors of including the LHC~\cite{Aad:2014vwa,Chatrchyan:2014gia} and Tevatron~\cite{Abazov:2011vi,Abulencia:2007ez} data in the global fit. The ratio of the 68\% C.L. errors to the baseline fit is shown and the NLO (NNLO) case is shown in the left (right) figures. For the LHC case, $R_{\rm high}$ and $p_\perp^{\rm jet}$ are taken.}
\label{fig:combrelerr2}
\end{center}
\end{figure}

In Fig.~\ref{fig:comberr} we show the impact at NNLO of the ATLAS and CMS jet data on the gluon PDF uncertainty, for the two choices of jet radii. As in the case of the central values, we find that the difference due to the scale choice is minimal, and so we only show results for the $p_\perp^{\rm jet}$ scale. The overall impact is seen to be moderate, although not negligible. To give a clearer comparison, we show the ratios to the baseline PDF uncertainty in  Fig.~\ref{fig:combrelerr} (left). For the higher $R$ choice, for low and intermediate values of $x$ the error reduction relative to the baseline ranges from $10-20\%$, but for the $x\sim 0.05-0.2$ there is little reduction and in some regions even a slight increase in the error. At high $x$ there is again a reduction in the uncertainty, although as $x$ approaches 1 and the jet data places little or no constraint, the quantitative result cannot be taken completely literally, as this will depend on the precise choice of PDF parameterisation. For the lower $R$ choice the reduction in the PDF uncertainty is less significant, and the $x$ region where this increases relative to the baseline is wider. In Fig.~\ref{fig:combrelerr} (right) we should the results for the higher jet radius choice and for different treatments of the ATLAS systematic errors. We can see that the partial decorrelation leads to a similar, although in some places slightly less constraining, impact on the uncertainties across the entire $x$ region in comparison to the default treatment, consistent with the impact on the central values shown before. On the other hand, for fully decorrelated uncertainties the impact at high $x$ in particular is much less constraining, although in the $x\sim 0.1$ region the uncertainties are in fact somewhat smaller.

In Fig.~\ref{fig:combrelerr1} (we show the impact of fitting the ATLAS and CMS data individually on the PDF uncertainties. We can see that, consistently with the results of the previous section, the impact of the ATLAS is generally larger, in particular at higher $x$, where the CMS data in fact lead to a somewhat larger uncertainty in comparison to the baseline. Including both the ATLAS and CMS data generally leads to some decrease in the uncertainties in comparison to the individual fits. In Fig.~\ref{fig:combrelerr2} we show the impact of the fits to the LHC and Tevatron data individually, as well as to the combination, at NLO and NNLO. We can see that with the exception of the intermediate $x\sim 0.05-0.1$ region at NNLO, the LHC data has a greater impact in reducing the PDF uncertainties. For the combined fit, the relative uncertainties reduce by $\sim 20\%$ across the entire $x$ region at NNLO, while at NLO the reduction in uncertainty is somewhat milder in comparison to the LHC--only fit in the $x\sim 0.01-0.1$ region, while in the highest $x$ regions the impact is somewhat larger. We can also see that, with the exception of this very high $x$ region, which will in any case be sensitive to parameterisation effects, the impact of the NNLO fit on the gluon is more significant in comparison to the NLO for all data combinations. Again, it will be interesting to see how this situation changes at NNLO when the full NNLO corrections are included in the Tevatron predictions.

\section{Conclusions and outlook}\label{sec:conc}

Inclusive jet production data has played a key role in constraining the partonic structure of the proton, and in particular the gluon at higher $x$, in global PDF fits. The availability of high precision jet data from the LHC combined with the recent release of the NNLO corrections to the hard cross section therefore provides an invaluable tool for high precision PDF constraints. 

In this paper, we have presented a detailed study of the impact of LHC jet data on a PDF fit within the MMHT global fitting framework, at NNLO. We have observed that to reliably perform such a study, certain issues require a careful treatment. Namely we have had to address the choice of jet scale and radius, and the fact that a satisfactory description of the systematics dominated ATLAS data cannot by default be achieved across the full kinematic region. After analysing the structure of the systematic shifts induced in describing the ATLAS data, we have determined a straightforward and minimal method to improve the fit quality; by decorrelating two sources of systematic uncertainty in rapidity, a greatly improved description is achieved. Crucially, despite this change in the fit quality,  we have shown that this only has a relatively small impact on the determination of the gluon itself in comparison to the default treatment. The result of our minimal approach should then be in line with a more complete consideration of different decorrelation scenarios permitted by experimental considerations.  This suggests that despite this question of the default fit quality, these data can still be reliably included in a PDF fit. On the other hand, we have found that decorrelating all sources of uncertainty in rapidity, in essence the approach that is assumed if only one rapidity bin is fitted, leads to larger shifts. Some caution in applying such a procedure therefore appears to be warranted.

We have then presented the fit quality at NLO and NNLO when the ATLAS and CMS jet data are included in a MMHT fit, for both the inclusive and leading jet $p_\perp$ scale choices, and different values of the jet radius $R$. We find that some improvement is in general achieved when going to NNLO, with the exception of the $p_\perp^{\rm jet}$ and lower $R$ choice, where there is a slight deterioration. The impact on the gluon PDF is qualitatively similar between orders. Although the theory predictions are quite different at lower jet $p_\perp$ when considering the two scale choices, we find that the fit quality including a proper treatment of the experimental systematics is in fact similar. Moreover, the impact on the gluon itself is very stable between the choices. This suggests that at least for the datasets under consideration in this paper, the effect of the choice of jet scale on PDF determination may not be as significant at NNLO as has sometimes previously been assumed.

In terms of the jet radius, the ATLAS data in particular has shown some preference of the larger ($R=0.6$) choice, although again the impact on the gluon is relatively stable in comparison to the smaller choice. In all cases the jet data are found to consistently prefer a somewhat softer gluon at high $x$ and a harder gluon in the intermediate $x$ region, with in general some $\sim$ 10--20 \% {\it relative} reduction in the PDF uncertainty. 

Thus, in this paper we have shown that LHC jet data may be reliably included in to global PDF fits at NNLO, while addressing in a minimal way the issue related to achieving a good description of the high precision, systematics dominated, ATLAS data across the whole kinematic region. We have only considered the 7 TeV data, for which the NNLO calculations are available. In future global fits, we will take our partially decorellated treatment of the experimental systematic errors for these datasets. However, in the future we intend to confirm if the above conclusions hold in the case of the 8 and 13 TeV jet data from the LHC\footnote{We note that the Run--II data sets are by default made available for a single common value of jet radius, $R=0.4$, preventing any comparison of different jet radii. Interestingly, we have seen here that a better description of the ATLAS data in particular may be achieved for a higher choice of jet radius, $R=0.6$.}. Moreover, this issue related to the description of the ATLAS data may become increasingly relevant in the high precision LHC era, and may warrant a more detailed study in the future of both the experimental and theoretical sources of uncertainty.

\section*{Acknowledgements}

We are grateful to Pavel Starovoitov for useful discussions and invaluable help with the \texttt{NLOJet++} interface to \texttt{APPLgrid}. We are also grateful to Ulla Blumenschein, Amanda Cooper--Sarkar, James Currie, Nigel Glover, Claire Gwenlan, Bogdan Malaescu, and Matthias Schott for useful discussions. LHL thanks the Science and Technology Facilities Council (STFC) for support via grant awards ST/L000377/1 and ST/P004547/1. RST thanks the Science and Technology Facilities Council (STFC) for support via grant awards ST/L000377/1 and  ST/P000274/1.

\bibliography{references}{}

\begin{thebibliography}{10}

\bibitem{Gao:2017yyd}
J.~Gao, L.~Harland-Lang, and J.~Rojo,
\newblock (2017), 1709.04922.

\bibitem{Abramowicz:2015mha}
ZEUS, H1, H.~Abramowicz {\em et~al.},
\newblock Eur. Phys. J. {\bf C75}, 580 (2015), 1506.06042.

\bibitem{Harland-Lang:2014zoa}
L.~A. Harland-Lang, A.~D. Martin, P.~Motylinski, and R.~S. Thorne,
\newblock Eur. Phys. J. {\bf C75}, 204 (2015), 1412.3989.

\bibitem{Dulat:2015mca}
S.~Dulat {\em et~al.},
\newblock Phys. Rev. {\bf D93}, 033006 (2016), 1506.07443.

\bibitem{Ball:2017nwa}
NNPDF, R.~D. Ball {\em et~al.},
\newblock Eur. Phys. J. {\bf C77}, 663 (2017), 1706.00428.

\bibitem{Accardi:2016qay}
A.~Accardi, L.~T. Brady, W.~Melnitchouk, J.~F. Owens, and N.~Sato,
\newblock Phys. Rev. {\bf D93}, 114017 (2016), 1602.03154.

\bibitem{Alekhin:2017kpj}
S.~Alekhin, J.~Blumlein, S.~Moch, and R.~Placakyte,
\newblock Phys. Rev. {\bf D96}, 014011 (2017), 1701.05838.

\bibitem{Czakon:2016olj}
M.~Czakon, N.~P. Hartland, A.~Mitov, E.~R. Nocera, and J.~Rojo,
\newblock JHEP {\bf 04}, 044 (2017), 1611.08609.

\bibitem{Boughezal:2017nla}
R.~Boughezal, A.~Guffanti, F.~Petriello, and M.~Ubiali,
\newblock JHEP {\bf 07}, 130 (2017), 1705.00343.

\bibitem{Kidonakis:2000gi}
N.~Kidonakis and J.~F. Owens,
\newblock Phys. Rev. {\bf D63}, 054019 (2001), hep-ph/0007268.

\bibitem{Kumar:2013hia}
M.~C. Kumar and S.-O. Moch,
\newblock Phys. Lett. {\bf B730}, 122 (2014), 1309.5311.

\bibitem{deFlorian:2013qia}
D.~de~Florian, P.~Hinderer, A.~Mukherjee, F.~Ringer, and W.~Vogelsang,
\newblock Phys. Rev. Lett. {\bf 112}, 082001 (2014), 1310.7192.

\bibitem{Liu:2017pbb}
X.~Liu, S.-O. Moch, and F.~Ringer,
\newblock (2017), 1708.04641.

\bibitem{Abazov:2011vi}
D0, V.~M. Abazov {\em et~al.},
\newblock Phys. Rev. {\bf D85}, 052006 (2012), 1110.3771.

\bibitem{Abulencia:2007ez}
CDF, A.~Abulencia {\em et~al.},
\newblock Phys. Rev. {\bf D75}, 092006 (2007), hep-ex/0701051,
\newblock [Erratum: Phys. Rev.D75,119901(2007)].

\bibitem{Aad:2011fc}
ATLAS, G.~Aad {\em et~al.},
\newblock Phys. Rev. {\bf D86}, 014022 (2012), 1112.6297.

\bibitem{Aad:2013lpa}
ATLAS, G.~Aad {\em et~al.},
\newblock Eur. Phys. J. {\bf C73}, 2509 (2013), 1304.4739.

\bibitem{Chatrchyan:2012bja}
CMS, S.~Chatrchyan {\em et~al.},
\newblock Phys. Rev. {\bf D87}, 112002 (2013), 1212.6660,
\newblock [Erratum: Phys. Rev.D87,no.11,119902(2013)].

\bibitem{Ball:2014uwa}
NNPDF, R.~D. Ball {\em et~al.},
\newblock JHEP {\bf 1504}, 040 (2015), 1410.8849.

\bibitem{Ridder:2013mf}
A.~Gehrmann-De~Ridder, T.~Gehrmann, E.~W.~N. Glover, and J.~Pires,
\newblock Phys. Rev. Lett. {\bf 110}, 162003 (2013), 1301.7310.

\bibitem{Currie:2013dwa}
J.~Currie, A.~Gehrmann-De~Ridder, E.~W.~N. Glover, and J.~Pires,
\newblock JHEP {\bf 01}, 110 (2014), 1310.3993.

\bibitem{Currie:2016bfm}
J.~Currie, E.~W.~N. Glover, and J.~Pires,
\newblock Phys. Rev. Lett. {\bf 118}, 072002 (2017), 1611.01460.

\bibitem{Aad:2014vwa}
ATLAS, G.~Aad {\em et~al.},
\newblock JHEP {\bf 02}, 153 (2015), 1410.8857,
\newblock [Erratum: JHEP09,141(2015)].

\bibitem{Chatrchyan:2014gia}
CMS, S.~Chatrchyan {\em et~al.},
\newblock Phys. Rev. {\bf D90}, 072006 (2014), 1406.0324.

\bibitem{Martin:2017uyr}
A.~D. Martin and M.~G. Ryskin,
\newblock Eur. Phys. J. {\bf C77}, 218 (2017), 1702.01663.

\bibitem{Aaboud:2017dvo}
ATLAS, M.~Aaboud {\em et~al.},
\newblock JHEP {\bf 09}, 020 (2017), 1706.03192.

\bibitem{Aaboud:2017wsi}
ATLAS, M.~Aaboud {\em et~al.},
\newblock (2017), 1711.02692.

\bibitem{Khachatryan:2016mlc}
CMS, V.~Khachatryan {\em et~al.},
\newblock JHEP {\bf 03}, 156 (2017), 1609.05331.

\bibitem{Khachatryan:2016wdh}
CMS, V.~Khachatryan {\em et~al.},
\newblock Eur. Phys. J. {\bf C76}, 451 (2016), 1605.04436.

\bibitem{Nagy:2003tz}
Z.~Nagy,
\newblock Phys. Rev. {\bf D68}, 094002 (2003), hep-ph/0307268.

\bibitem{Carli:2010rw}
T.~Carli {\em et~al.},
\newblock Eur. Phys. J. {\bf C66}, 503 (2010), 0911.2985.

\bibitem{Applgridweb}
\texttt{https://applgrid.hepforge.org}.

\bibitem{Currie:2017ctp}
J.~Currie {\em et~al.},
\newblock Acta Phys. Polon. {\bf B48}, 955 (2017), 1704.00923.

\bibitem{Harland-Lang:2017dzr}
L.~A. Harland-Lang, R.~Nathvani, R.~S. Thorne, and A.~D. Martin,
\newblock Acta Phys. Polon. {\bf B48}, 1011 (2017), 1704.00162.

\bibitem{Harland-Lang:2016yfn}
L.~A. Harland-Lang, A.~D. Martin, P.~Motylinski, and R.~S. Thorne,
\newblock Eur. Phys. J. {\bf C76}, 186 (2016), 1601.03413.

\bibitem{Aad:2014bia}
ATLAS, G.~Aad {\em et~al.},
\newblock Eur. Phys. J. {\bf C75}, 17 (2015), 1406.0076.

\bibitem{ATLASpriv}
Ulla Blumenschein, Claire Gwenlan, Bogdan Malaescu and Matthias Schott, private
  communication.

\bibitem{Martin:2009iq}
A.~D. Martin, W.~J. Stirling, R.~S. Thorne, and G.~Watt,
\newblock Eur.Phys.J. {\bf C63}, 189 (2009), 0901.0002.

\end{thebibliography}
\bibliographystyle{h-physrev}

\end{document}